\documentclass[
  aip,
  author-year,
  sorting=ynt,
  reprint,
  onecolumn,
  tightenlines,
  12pt,
  notitlepage,
  amsmath,
  amssymb
]{revtex4-1}

\usepackage{graphicx}
\usepackage{dcolumn}
\usepackage{bm}
\usepackage{xcolor}
\usepackage{hyperref}
\usepackage{rotating}

\usepackage{lipsum,graphicx,subcaption}
\hypersetup{colorlinks=true,urlcolor=blue,citecolor=blue,linkcolor=blue}
\setcitestyle{square}

\begin{document}

\preprint{AIP/123-QED}

\title[]{Bidisperse and polydisperse suspension rheology  at large solid fraction}

\author{Sidhant Pednekar}
\affiliation{ 
Benjamin Levich Institute, New York, USA
}
\affiliation{Department of Chemical Engineering, The City College of New
York, NY, NY 10031.}

\author{Jaehun Chun}
 \email{jaehun.chun@pnnl.gov}
\affiliation{ 
Pacific Northwest National Laboratory
Richland, Washington 99352, USA
}

\author{Jeffrey Morris}
 \email{morris@ccny.cuny.edu}
\homepage{http://www-levich.engr.ccny.cuny.edu/~jmorris/.}
\affiliation{ 
Benjamin Levich Institute, New York, USA
}
\affiliation{Department of Chemical Engineering, The City College of New
York, NY, NY 10031.}

\date{\today}

\begin{abstract}
 At the same solid volume fraction, bidisperse and polydisperse suspensions display lower viscosities, and weaker normal stress response, compared to monodisperse
suspensions.  The reduction of viscosity associated with size distribution can be explained by an increase of the maximum flowable, or jamming, solid fraction $\phi_m$. 
In this work, concentrated or ``dense'' suspensions are simulated under strong shearing, where thermal motion and repulsive forces are negligible, but we allow for particle contact with a mild frictional interaction with interparticle friction coefficient of $\mu = 0.2$.   Aspects of bidisperse suspension rheology are first revisited to establish that the approach reproduces established trends; the study of bidisperse suspensions at size ratios of large to small particle radii of $\delta = 2$ to 4 shows that a minimum in the viscosity occurs for $\zeta$ slightly above 0.5, where $\zeta = \phi_{l}/\phi$ is the fraction of the total solid volume occupied by the large particles.  The simple shear flows of polydisperse suspensions with truncated normal and log normal size distributions, and bidisperse suspensions which are statistically equivalent with these polydisperse cases up to third moment of the size distribution, are simulated and the rheologies are extracted.  Prior work shows that such distributions with equivalent low-order moments have similar $\phi_m$, and the rheological behaviors of normal, log normal and bidisperse cases are shown to be in close agreement for a wide range of standard deviation in particle size, with standard correlations which are functionally dependent on $\phi/\phi_m$ providing excellent agreement with the rheology found in simulation.    
The close agreement of both viscosity and normal stress response between bi- and polydisperse suspensions demonstrates the controlling influence of the maximum packing fraction in noncolloidal suspensions. Microstructural investigations and the stress distribution according to particle size are also presented. 
\end{abstract}

\maketitle

\section{\label{sec:level1}INTRODUCTION}%}:\protect\\ 
Flowing suspensions are found industrially as cements and pastes, and as mud in nature. Under dilute solid volume fraction, i.e. small $\phi$, 
 the relative viscosity $\eta_r (\phi) = \eta_s(\phi)/\eta_0$  of a suspension in a Newtonian fluid of viscosity $\eta_0$ retains behavior that is quasi-Newtonian and the suspension viscosity differs only mildly from the suspending fluid, as in the Einstein viscosity $\eta_s(\phi) = \eta_0 (1 + 2.5\phi)$ or extensions which apply to $\phi<0.1$.  
In the examples noted, however, we encounter conditions far from dilute, as the particles approach their maximum packing fraction and the materials are often called {\em dense suspensions}. 
In these examples, the particles are also typically nonuniform in size, and this work addresses simulation of dense suspensions with both bidisperse and polydisperse size distributions. 

While experiments have probed size dispersion about the mean in suspensions, very limited dynamical simulation work has addressed polydisperse suspensions. Most laboratory studies examining size distribution have considered bidisperse suspensions, but continuously polydisperse distributions of particle size are more often encountered.  Particles following a normal distribution are common in processes when particles are purposefully generated around a specific mean size. Log-normal distributions of particles are often found in soils, aerosols, mining and grinding operations [\cite{wagner1994representing}] and  occur frequently in particle growth processes [\cite{soderlund1998lognormal}].
 In application, the influence of particle size distribution on rheology can be quite important, e.g. in the coal and food industries  [\cite{servais2002influence,boylu2004effect,liu2015investigation,singh2016influence,leverrier2016influence}]. Chocolate manufacturing, for example, typically requires control of the particle size distribution to facilitate pumping and mixing of molten chocolate [\cite{mongia2000role}] and transportation and grinding of dense milk suspensions [\cite{saeseaw2005sedimentation}]. 
  As a further example which directly motivates our work, the nuclear waste materials at the Hanford site in Washington state in the United States are  slurries of metal-oxide particles that have varying degrees of polydispersity [\cite{wells2007estimate,chun2010stabilization,wells2011hanford,clark2016basic}].  These are most often nonspherical particles, but here the particle size distribution is our focus and we consider only spheres. While research dedicated towards studying the rheology of slurries with particle size distributions, e.g. in the range of 1 - 100 $\mu$m [\cite{chun2011effect}], provides valuable system-specific information, here we consider a wide range of the polydispersity parameters, toward development of an understanding of how the size distribution of dense suspensions impacts upon the shear and normal stresses in simple shear.   \\
 
\noindent One factor imposing difficulty on the study of polydispersity on suspension rheology is that making precise particle size distributions (PSD) is experimentally challenging. Computer simulations provide a powerful tool for such examinations.  \cite{chang1993dynamic, chang1994effect, chang1994rheology}, for example, used Stokesian Dynamics simulations to study the effect of bidispersity on suspension rheology. However, to date, these studies are among only a few to address simulation of even the bidisperse condition.  One hurdle to simulation of polydisperse suspensions is the complexity of representing the interaction of particles, particularly at large separations of particles where long-range hydrodynamics play a role, although recent work provides a method to remove this obstacle [\cite{wang2015short}].  To circumvent this difficulty, for dense suspensions where particles are extremely crowded and neighboring particle surfaces are always very close to contact, we note that it is plausible that such long-range effects are secondary in importance to the singular lubrication effects of the fluid [\cite{ball1997simulation}].  In addition, the possibility of contact should not be discounted especially in modeling of materials with some surface roughness or angularity, as one finds for platelike particles in natural mud, or  crystalline particles in nuclear waste slurries; contact has also been shown to have potentially important effects in cohesive particle suspensions [\cite{pednekar2017simulation}].  Recent work has developed a simulation method to capture lubrication and contact effects in dense suspensions [\cite{mari2014shear}], and we employ this to explore systematically the effect of polydispersity.  

It will be shown that, despite complex microstructural variation at different conditions studied, the viscosity is remarkably well-predicted by simply establishing the maximum packing fraction and using an empirical form for the relative viscosity,   e.g. the form $\eta_r (\phi/\phi_m) = (1-\phi/\phi_m)^{-2}$ [\cite{maron1956application}].  The normal stress behavior is similarly reduced to a function of $\phi/\phi_m$.    We first review relevant literature providing a background on bi- and polydisperse suspension rheology as well as methodology for predicting maximum packing fraction; in doing so, we will introduce the basic parameters describing the suspensions studied.

\section{\label{sec:level2}Literature}

It is accepted that bidispersity or polydispersity reduces viscosity for the same solid loading (solid volume fraction). The maximum packing fraction, $\phi_m$, is also known to increase with greater polydispersity.  The simple explanation, valid for large difference between the large and small particles of the distribution, is that the small particles may fit into the interstices between larger particles. The reduction of viscosity and increase in maximum packing of polydisperse suspensions can be utilized to increasing flowability and/or solid content of suspensions. Despite the prevalence of polydispersity in application, most existing studies on the effect of varying sizes in suspensions focus on bidisperse suspensions. The state of understanding on bidisperse and higher order polydisperse suspension rheology is summarized below using a few key experimental and simulation studies.

\subsection{\label{sec:head}Experiments and Simulation}

\subsubsection{\label{sec:bidi_lit}Bidisperse suspensions}

In addition to the volume fraction ($\phi$) used to characterize monodisperse suspensions, two additional parameters are traditionally used to describe bidisperse suspensions.  These can be chosen differently, but the following forms are standard.  These are the size ratio  

\begin{equation}
\delta = a_l/a_s \label{delta},
\end{equation}
and the fraction of the solid volume occupied by the large particles,  
\begin{equation}
\zeta = \phi_{l}/\phi, \label{zeta}
\end{equation}
 
\noindent where the large particles have radius $a_l$ and the small have radius $a_s$, while $\phi$ and $\phi_l$ are the bulk and large-particle solid volume fractions.  \cite{shapiro1992random} experimentally studied bidisperse suspensions using non-Brownian glass beads in glycerin. They found a decrease in viscosity as they moved from monodisperse suspensions ($\zeta$ = 0 and 1) to bidisperse suspensions ($0 < \zeta < 1$). At fixed $\phi$ and $\delta$ (for $\delta$ = 2 and 4) they observed a decrease in viscosity with $\zeta$ to a minimum. The reduction in viscosity with bidispersity was found to be more pronounced at higher $\delta$. \cite{chong1971rheology} and \cite{gondret1997dynamic} made similar observations. \cite{barnes1989introduction} noted up to a 50-fold reduction in viscosity in going from a concentrated monodisperse to bidisperse suspension at the same $\phi$. \cite{poslinski1988rheological} showed an increase in $\phi_m$ and a corresponding decrease in shear viscosity, first normal stress difference, dynamic viscosity and storage modulus with bidispersity at different $\phi$.  \cite{chang1993dynamic, chang1994effect, chang1994rheology} used Stokesian Dynamics to calculate
hydrodynamic interactions in bidisperse suspensions and showed trends which matched the general experimental observations noted above. More recently, \cite{wang2015short} used conventional Stokesian Dynamics to study the short-time transport properties of bidisperse colloidal suspensions.  In addition to rheological modifications, bidisperse suspensions in nonuniform shear such as pressure-driven flow, where particle migration occurs, are known to display segregation behavior based on particle size [\cite{lyon1998experimental,semwogerere2008shear}], but here we will focus on conditions where the particle sizes remain well-mixed.\\
\subsubsection{\label{sec:level3}Polydisperse suspensions}

To characterize polydisperse suspensions, $\delta$ and $\zeta$ are inconvenient.  The polydispersity index ($\alpha$) has, instead, more commonly been used for this purpose [\cite{pusey1987effect,rastogi1996microstructure}]. A polydispersity factor is defined as the standard deviation normalized by the mean of the distribution, 

\begin{equation}\label{polyd}
\alpha = \sqrt{\langle \Delta a^2 \rangle}/\langle a \rangle, 
\end{equation}

\noindent where $\Delta a$ = $a - \langle a \rangle$, and $a$ is the particle radius.
Polydispersity is thus described at leading order as a measure of the spread or variance around the mean. \cite{luckham1999effect} experimentally studied the rheology of three different  polydisperse suspensions with varying degrees of polydispersity.  They noticed that the broadest size distribution suspension had the lowest effective viscosity. \cite{rastogi1996microstructure} used non-equilibrium 
Brownian dynamics simulations to study the effect of polydispersity on the rheology and microstructure of charged suspensions following a Schulz
distribution, and also observed decreasing viscosities with increasing polydispersity.    Polydisperse suspensions are also known to have weaker shear thickening than their monodisperse counterpart [\cite{boersma1990shear}].  
\subsection{\label{sec:level2}Modelling}

\subsubsection{\label{sec:monomod}Monodisperse suspensions}

Various empirical forms for relative viscosity are used at high solid concentrations.  \cite{maron1956application} proposed the form 
\begin{equation}
\eta_r = (1 - \phi/\phi_m)^{-2} \label{one}. 
\end{equation}
\cite{krieger1959mechanism} proposed a different exponent of the form $-[\eta]\phi_m$ to the above equation where $[\eta]$ is the intrinsic viscosity.  \cite{frankel1967viscosity} proposed the following, 

\begin{equation}
\eta_r = \frac{9}{8}\bigg\{\frac{(\phi/\phi_m)^{1/3}}{(1-(\phi/\phi_m)^{1/3})}\bigg\},  \label{two}
\end{equation}
and \cite{ferrini1979shear} suggested an empirical relation of the form
\begin{equation}
\eta_r = \bigg\{1 + \frac{\frac{1}{2}[\eta]\phi}{1 - \phi/\phi_m}\bigg\}^2.  \label{three}
\end{equation}
Common to the approaches is the appearance of a maximum packing fraction, $\phi_m$, at which the viscosity diverges.  This implies that the limit $\phi/\phi_m\rightarrow 1$ is an approach to a jammed condition.   A range of values from $\phi_m$ = 0.53 [\cite{lewis1968viscosity}] to $\phi_m$ =  0.71 [\cite{de1985hard}] for monodisperse systems have been suggested in the literature, but it has become 
clear in recent years that the maximum packing fraction varies between the random loose-packed state, $\phi_{rlp}$ found for friction-dominated conditions, to the random close-packed state, $\phi_{rcp}$ found when particles have well-lubricated and frictionless interactions.\\ 

\noindent The normal stress differences given by (\ref{n1}, \ref{n2}) are known to follow forms similar to the viscosity models (\ref{one},\ref{two},\ref{three}) [\cite{poslinski1988rheological}]; the basic argument supporting the similarity of the various viscometric functions  is that all stresses in the dense limit are due to similar shear-driven mechanisms [\cite{morris1999curvilinear}], namely moments of the lubrication and contact stress distributions over particle surfaces.  

\subsubsection{\label{sec:polymod}Bidisperse and polydisperse suspensions}

\cite{farris1968prediction} developed a method for predicting the
viscosity of suspensions with multimodel size distributions.  The approach was developed for suspensions with large size ratios between distinct particle groups, based
on the idea that interaction between the different groups can be represented by assuming the suspension of finer particles to behave
as a liquid with viscosity given by the effective suspension viscosity at its volume fraction. This approach has been modified for finite $\delta$ bidisperse [\cite{zaman1998rheology}] and recently, polydisperse  [\cite{mwasame2016modeling}] suspensions.\\

\noindent One well-known approach to modeling the rheology of polydisperse suspensions is with models similar to (\ref{one} - \ref{three}), with the use of the corresponding polydisperse $\phi_m$  [\cite{luckham1999effect,servais2002influence,pishvaei2006modelling,qi2011relative,shewan2015analytically}]. The success of this approach, in prior as well as the present work, indicates that the reduced volume fraction ($\phi/\phi_m$) plays a central role in the mixture flow well away from jamming. 
 As in the case for monodispersity, the maximum packing of bidisperse and polydisperse particle distributions has been the subject of study [\cite{ouchiyama1980estimation,gupta1986maximum,santiso2002dense,brouwers2006particle}]. More recently, new approaches have been introduced to study the maximum packings of polydisperse systems [\cite{farr2009close,baranau2014random,desmond2014influence}]. 

\subsection{\label{sec:thiswork}This work}

\noindent In this study, we use simulation to probe  the role of particle size distribution on suspension rheology in the dense non-colloidal regime. We begin by considering the effect of bidispersity. In addition to reproducing experimentally-observed trends showing bidispersity to reduce viscosity, we gain insight into the variation of normal stress differences ($N_1$, $N_2$) and particle pressure ($\Pi$). Furthermore, we examine the microstructure of these bidisperse suspensions, via pair distribution functions.
We then describe polydisperse suspension rheology based on simulations of suspensions of truncated normal and log-normal distributions. 
 A central finding is that the apparent complexity of these polydisperse suspensions can, for bulk rheology, be largely reduced by considering low-order moments of the size distribution.  This is demonstrated by comparison of the results with relatively well-understood bidisperse suspensions of equivalent rheology. The basis for this behavior is explored by considering the stress contributions 
 of particles as a function of their size in the bi- and polydisperse suspensions, an analysis tool which is presently unique to simulation. \\

\noindent The starting point for connecting the rheologies in bi- and polydisperse suspension is the determination of the reduced solid volume fraction $\phi/\phi_m$ for non-monodisperse  suspensions. \cite{chong1971rheology} show that the relative viscosities of glass suspensions (both monodisperse and bidisperse) plotted against reduced solid volume fraction ($\phi/\phi_m$) collapse on to a single curve. \cite{chang1994effect} further tested this approach with experimental results of several works. The review by \cite{stickel2005fluid} also present a similar plot, 
and the authors suggest that the reduced solid volume fraction provides a bulk parameter which controls the suspension microstructure which ultimately governs flow behavior. \\
\begin{figure}
\begin{minipage}[b]{0.33\linewidth}
    \centering
   a)\includegraphics[width=1\linewidth]{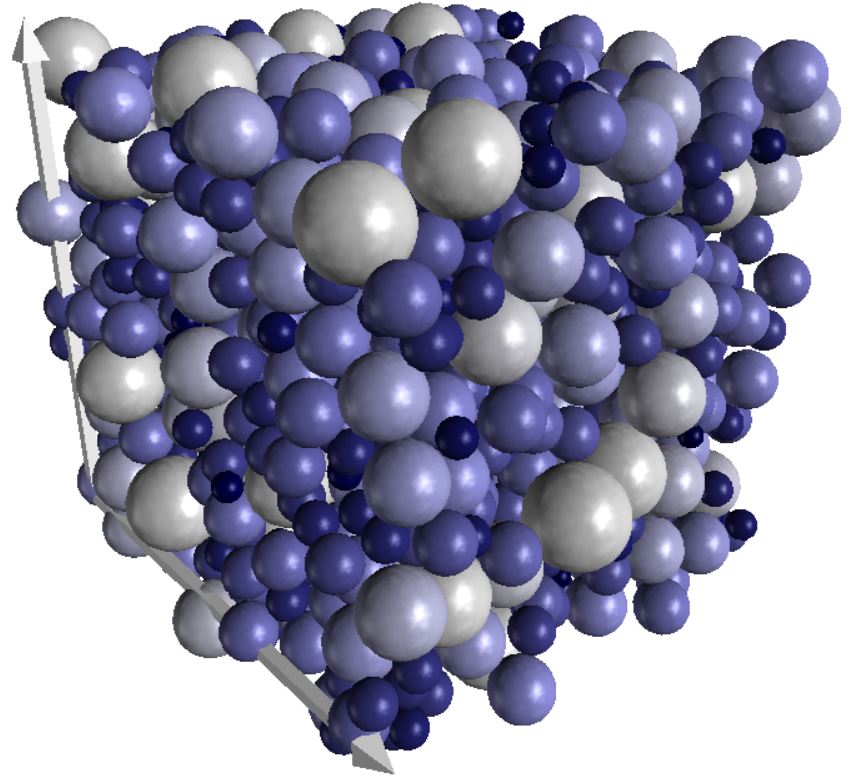} 
  \end{minipage}\qquad
  b)\begin{minipage}[b]{0.32\linewidth}
    \centering
    \includegraphics[width=1\linewidth]{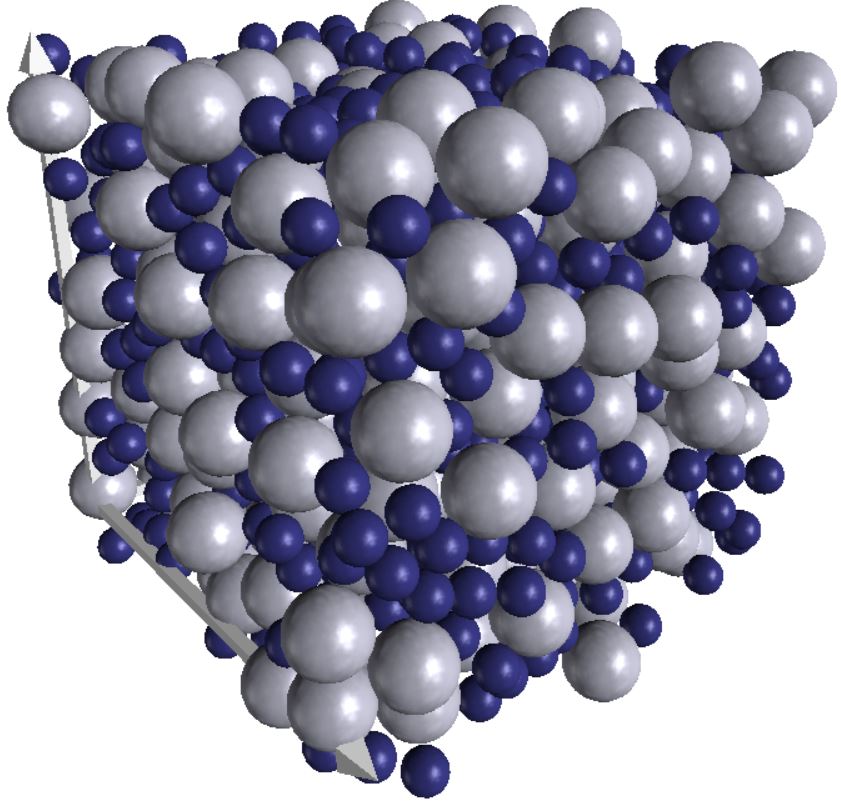} 
  \end{minipage}\qquad
  \begin{minipage}[b]{0.055\linewidth}
    \centering
    \includegraphics[width=1\linewidth]{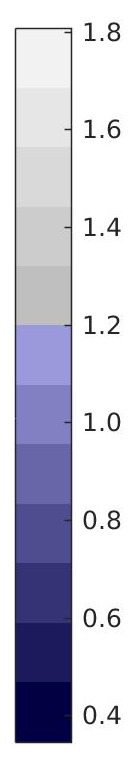} 
  \end{minipage}\par
\caption{Illustration of a) normal distribution with b) the rheologically equivalent bidisperse system. The colorbar indicates the size of individual particles.}
\end{figure}

\noindent Determination of maximum packing fraction, and its relation to statistical measures of the particle size distribution (PSD), is thus of central importance. Desmond \& Weeks [\cite{desmond2014influence}] studied polydisperse packings using a simulation algorithm based on the expansion of infinitesimal points to particles whose size followed a specified PSD. This work demonstrated that particles with different PSDs but having the same mean particle size, standard deviation and skewness around the mean have similar maximum packing. Recall that the standard deviation and skewness are normalized second and third moments of the distribution.  Although the approach is empirical and its generality is still untested, it was shown to be successful for particles of several distributions, including binary, linear, normal, and log normal. The maximum packing obtained for such granular assemblies having an isotropic microstructure has been shown to have a direct correspondence to the maximum packing (or jamming point) obtained from shear flow viscosity measurements [\cite{shapiro1992random,probstein1994bimodal}]. Another approach has emerged recently for constructing equivalent bidisperse systems with the same hard sphere equation of state as polydisperse systems [\cite{ogarko2012equation}]. We will however apply the approach of Desmond \& Weeks as the basis for a framework to determine rheologically similar equivalent bidisperse and polydisperse suspensions, as this has been shown to predict maximum packing with better accuracy. \\
 
\section{Model and methods}{\label{model}

\noindent We simulate the rheology of dense bidisperse and higher order polydisperse non-colloidal suspensions.  The simulation method combines 'lubrication flow' (LF) description of hydrodynamic interactions with discrete element modeling (DEM) of contacts between particles [\cite{mari2015discontinuous}], and is referred to as LF-DEM. As discussed in Sec. \ref{sec:bidi_lit}, polydispersity effects on rheology have pronounced effects at large volume fractions.  For such dense suspensions, contact forces are expected to play a role [\cite{boyer2011unifying}] and we assume that hydrodynamic interactions can be satisfactorily represented by pair-wise additive short range lubrication forces [\cite{ball1997simulation}]. This simulation tool has been shown to reproduce important aspects of dense suspension rheology, including continuous and discontinuous shear-thickening [\cite{mari2015discontinuous,seto2013discontinuous,mari2014shear}] and recently, the obscuring of shear thickening by  attractive forces and a resulting large low shear viscosity or yield stress [\cite{pednekar2017simulation}]. We simulate neutrally-buoyant particles of variable radius $\sim a$, suspended in viscous fluid (density $\rho$ and viscosity $\eta_0$) sheared at rate $\dot{\gamma}$  in the Stokes limit, i.e., small particle-scale Reynolds number, $Re = \rho \dot{\gamma}a^2/\eta_0 \ll 1$. The influence of Brownian motion is assumed to be negligible, so that the P\'eclet number satisfies $Pe = 6 \pi \eta_0 a^3 \dot{\gamma}/kT \gg$ 1. 

We solve the overdamped Langevin equation for the particle motion, 
\begin{equation} 0 = \bm{F}^H + \bm{F}^C,
\end{equation}
 where $\bm{F}^H$ and $\bm{F}^C$ are hydrodynamic and contact forces, respectively. A detailed explanation of these forces is provided in \cite{mari2014shear}, and our description will hence be kept brief. The hydrodynamic forces are of the form $\bm{F}^H = -\bm{R}_{FU}\cdot(\bm{U}-\bm{U}^\infty) + \bm{R_{FE}:E}^\infty$, with $\bm{U}^\infty =  \dot{\gamma} y \hat{\bf e}_x$ being the flow due to imposed shear and $\bm{E}^\infty$ the associated rate-of-strain tensor described by $\bm{E}^\infty$ $\equiv \frac{\dot{\gamma}}{2}
 (\hat{\bf e}_x \hat{\bf e}_y + \hat{\bf e}_y \hat{\bf e}_x)$. The hydrodynamic resistance matrices $\bm{R}_{FU}$ and $\bm{R_{FE}}$ contain leading order terms corresponding to short-range lubrication forces [\cite{ball1997simulation}]. Particle roughness ($\sim 10^{-3}a$) is introduced in the simulation to regularize the singularity associated with lubrication and allow interparticle contacts, as described in prior work \cite{mari2014shear}. These contacts are modeled as a linear spring following the Coulomb friction law, ${F}_{\rm tan}^C \leq \mu {F}_{\rm nor}^C$ [\cite{luding2008cohesive}], where $\mu$ is the coefficient of interparticle friction. \\

\noindent The normal and log-normal distributions are developed by using using random number generator functions in MATLAB, specifically \textit{normrnd} and \textit{lognrnd} (Matlab R2016b) with the required mean and variance. The discretized analogues of these continuous distributions are represented using $N = 1000$ particles and the distribution from among $10^4$ generated distributions for each case is chosen as that one having the least square error relative to theoretical curves (for example, see Fig. \ref{fgr:fgr}c-d for the probability density function. The bidisperse suspensions studied in this work are limited to size ratios of $\delta \leq$ 4. The maximum polydispersity ($\alpha$) examined for normal distributions is $\alpha$ = 0.2 and for log-normal distributions is $\alpha$ = 0.3. The generated discrete particle distributions at these $\alpha$ have $>$ 96 $\%$ of the particle radii fall in the range 1.6 $\geq$ a/$\langle$a$\rangle$ $\geq$ 0.4, once again a size ratio of about four between largest and smallest particles in the distribution. These particle radii are resolved to the second decimal place. In order to ensure sufficient distribution of particles in the extreme ends of the normal and log-normal distributions to ensure statistically meaningful results, simulations with $N = 2000$ were also performed. The rheology of these polydisperse systems is seen to be largely insensitive to finite size scaling effects, by comparison of results using $N = 2000$ instead of $N = 1000$ in the unit cell of the simulation. Lees-Edwards periodic boundary conditions are used in simulations in a cubic unit cell. All simulations reported are run over a period of 30 strain units, discarding results from a short transient period of 2 strain.

\section{Results}

\noindent We first simulate the flow of bidisperse suspensions and examine their rheological properties, considering the bidispersity parameter in the full range, $0\le \zeta \le 1$. At the limits $\zeta = 0$ or 1, the suspension is monodisperse (all small or all large particles).   Dense monodisperse suspensions are known to order into layers under flow causing a reduction in viscosity as these layers slip past each other [\cite{sierou2002rheology,kulkarni2009ordering}]. Bidispersity or polydispersity and frictional interactions tend to break up this ordering [\cite{rastogi1996microstructure,seto2013discontinuous}]. Ordering and its effects are undesired as we seek to understand the disordered material behavior, and thus we seek to avoid ordering by modeling contact interactions with a modest friction coefficient ($\mu_{fric}$ = 0.2). This was found based on our observations to be sufficient in eliminating any appreciable long-range ordering. Furthermore, this friction coefficient is in the range of experimentally measured friction coefficients between non-Brownian particles  [\cite{comtet2017pairwise}].

\subsection{\label{sec:bidi}Bidisperse suspensions}

\begin{figure}[b]
\centering
\includegraphics[width=.6\linewidth]{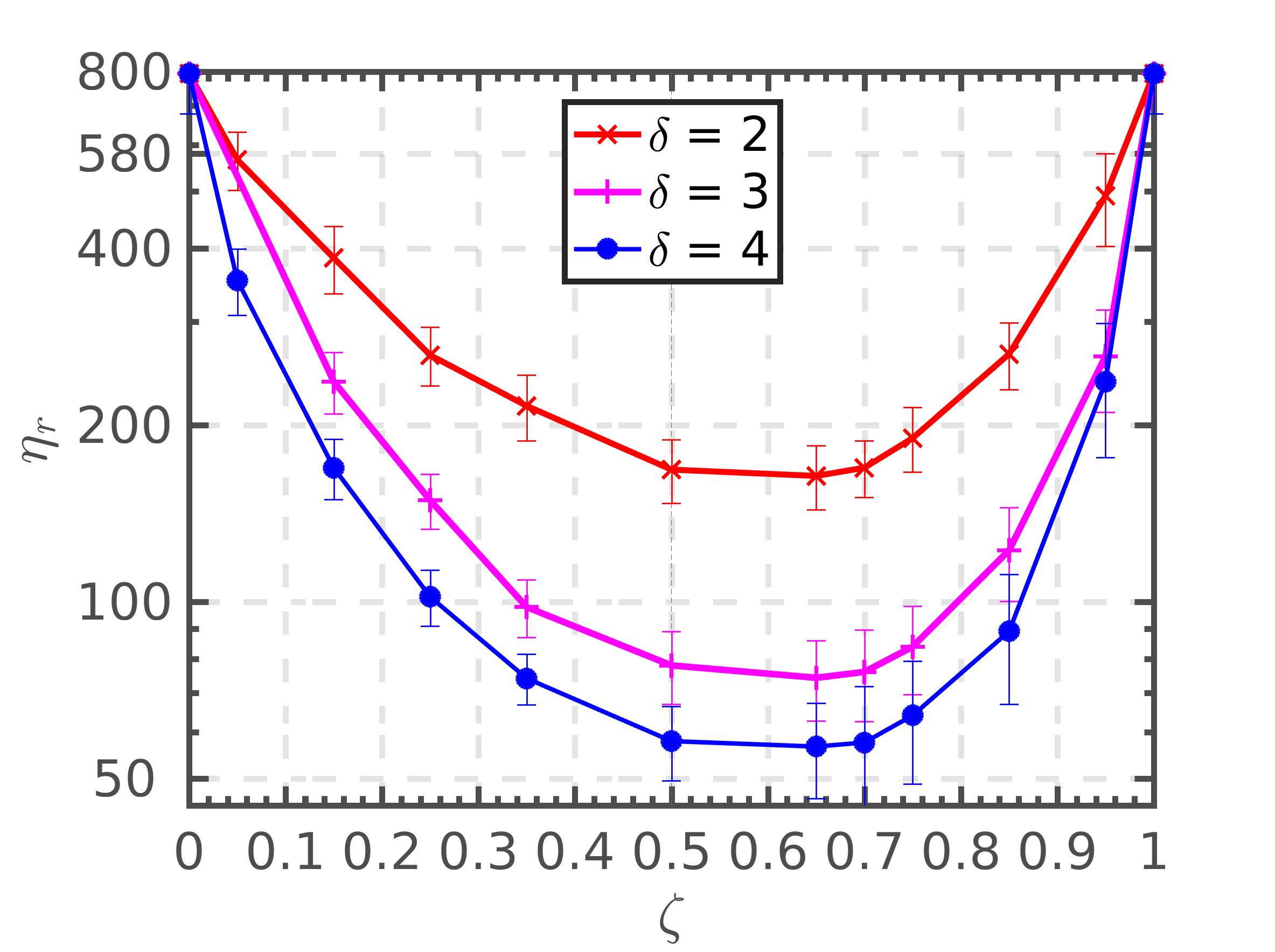}
\caption{Relative viscosity of bidisperse suspensions (log-scale) as a function of size ratio $\delta$ and large particle fraction of total solid loading $\zeta=\phi_l /\phi$ at fixed $\phi$ = 0.6.}
\label{fig:eta_delta}
\end{figure}

\noindent\textit{Effect of size ratio $\delta$ and composition $\zeta$}:  As noted in Sec.\ref{sec:bidi_lit}, the total volume fraction, $\phi$, is complemented by two additional parameters to characterize bidisperse suspensions, namely the size ratio, $\delta$, and $\zeta =\phi_l/\phi$ giving the large-particle fraction. In Fig. \ref{fig:eta_delta}, we show the effect of bidispersity on the simulated relative viscosity, $\eta_r$, as a function of $\zeta$ for $\delta =2$, 3, and 4, at $\phi = 0.6$. The end points $\zeta = 0$ and 1 are monodisperse and exhibit identical rheology due to size invariance of the non-colloidal suspension.   At a fixed size ratio of $\delta$ = 2, $\eta_r$ exhibits a marked decrease in viscosity with $\zeta$ increasing from 0, reaching a minimum at $\zeta \approx 0.65$, and then increasing back to the monodisperse viscosity at $\zeta =1$. As seen in Fig. \ref{fig:eta_delta}, the reduction in viscosity is more pronounced at higher size ratio ($\delta$ = 3, 4) displaying over a 90$\%$ reduction from the monodisperse case at the minimum with respect to $\zeta$ at $\delta = 4$. \\

\noindent The volume composition at minimum viscosity ($\zeta_{min}$) is of interest. The line representing $\zeta = 0.5$ in Fig. \ref{fig:eta_delta} is accentuated to show better that the reduced viscosity minima are at $\zeta>0.5$. A method for estimation of $\zeta_{min}$ has been proposed [\cite{greenwood1997effect}] for $\delta$ $\rightarrow \infty$ systems. With larger particles packed to monodisperse $\phi_{m}$, one can assume that for large size ratios the voids between coarser particles are completely accessible to smaller particles and they too will pack to the $\phi_{m}$. Hence, $\phi_{tot}$ will be\\

\begin{equation}
\phi_{tot}= \phi_{m} + (1-\phi_{m})     \phi_{m}.
\end{equation}

\noindent Using monodisperse $\phi_m$ $\approx$ 0.624, we find

\begin{equation}
\phi_{tot} = 0.858.
\end{equation}
Hence, \begin{equation} \zeta_{min} = \frac{0.624}{0.858}  = 0.727.\end{equation} 

\noindent We find values of $\zeta_{min}$ which are slightly smaller, in the range of $\zeta_{min}=0.65- 0.7$ in Fig. \ref{fig:eta_delta}, for the range of $\delta$ studied.\\

\textit{Bidisperse rheology with volume fraction}: The variation of relative viscosity, $N_1$, $N_2$ and particle pressure ($\Pi$) with bidispersity for $0.54 < \phi < 0.6$ is presented in Fig \ref{fgr:bi}a-d. The normal stress differences and particle pressure have been shown to be of particular relevance in particle migration and segregation studies [\cite{miller2006normal,morris2009review,boyer2011dense}]. These quantities are defined as follows:\\
\begin{equation}
N_1 = \Sigma_{11} - \Sigma_{22},\\ \label{n1}
\end{equation}

\begin{equation}
N_2 = \Sigma_{22} - \Sigma_{33,} \\
\label{n2}
\end{equation}
and 

\begin{equation}
\Pi = -\frac{1}{3}(\Sigma_{11}+\Sigma_{22}+\Sigma_{33}).\\
\end{equation}

\noindent Here 1, 2, 3 are the flow, velocity gradient and vorticity directions, respectively. In Fig \ref{fgr:bi}a we plot the variation of $\eta_r$ as a function of the large particle fraction $\zeta$ at different volume fractions for $\delta$ = 3. We observe greater reductions in viscosity at higher $\phi$. This result arises because $\eta_r$ of a monodisperse suspension at $\phi \rightarrow \phi_m$ is arbitrarily large, while a bidisperse suspension at the same $\phi$ is displaced from its maximum packing fraction and remains flowable. The $N_1$ variation with bidispersity seen in Fig. \ref{fgr:bi}b varies between small positive and negative values with $\zeta$ with relatively large  uncertainty. The abrupt decrease of $N_1$ with even slight bidispersity seen in simulation may be a source of some of the experimental difficulty in measuring $N_1$ [\cite{denn2014rheology,gamonpilas2016shear}].  The second normal stress difference is negative, and this quantity as well as the particle pressure $\Pi$ in Fig. \ref{fgr:bi}c-d show reductions in magnitude with bidispersity. Osmotic pressure (shown to be related to the particle pressure [\cite{yurkovetsky2008particle}]) of bidisperse Brownian suspensions has been observed to show similar trends [\cite{kim1993some}]. \\

\begin{figure}[!t] 
%\captionsetup{singlelinecheck=false, justification=centering}
  a) \begin{minipage}[b]{0.42\linewidth}
    \centering
    \includegraphics[width=1.15\linewidth]{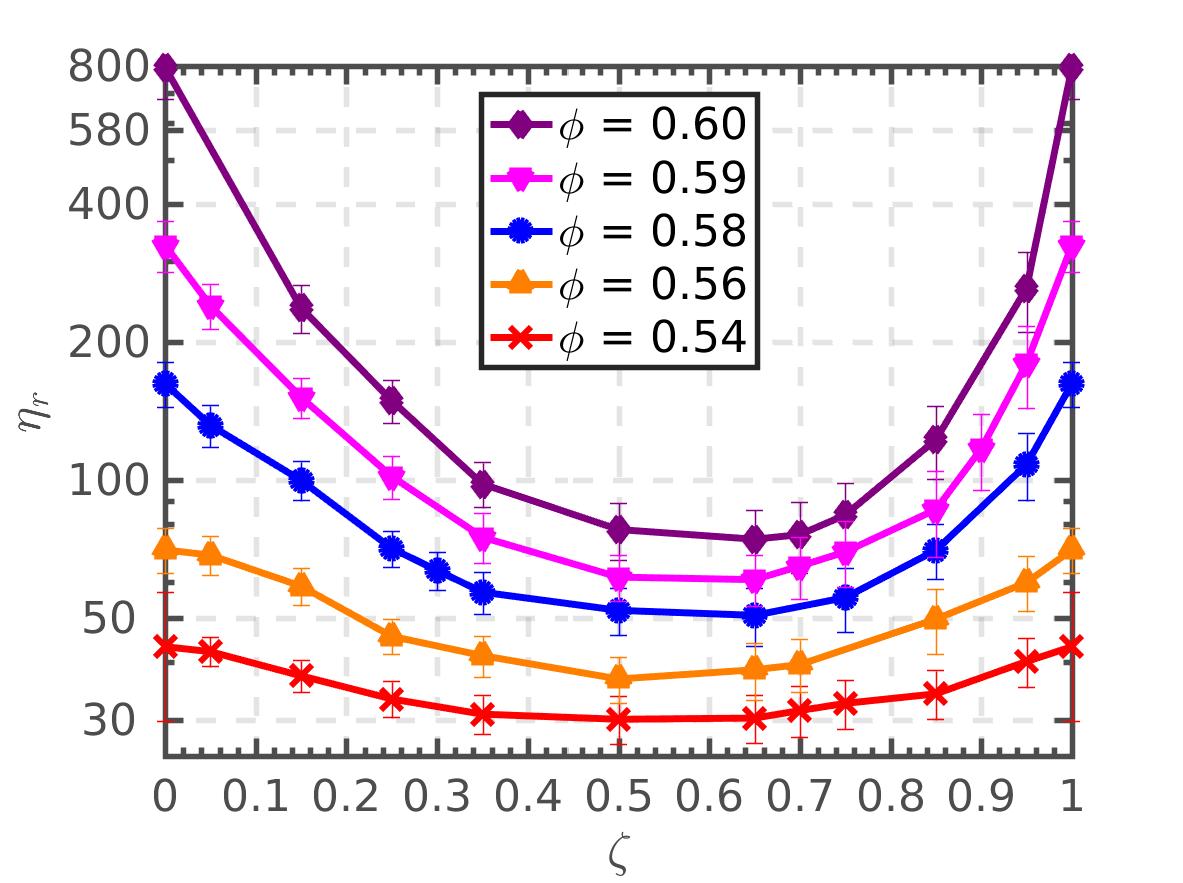} 
    %\caption{Initial condition} 
    %\vspace{4ex}
  \end{minipage}%%
  b) \begin{minipage}[b]{0.42\linewidth}
    \centering
    \includegraphics[width=1.15\linewidth]{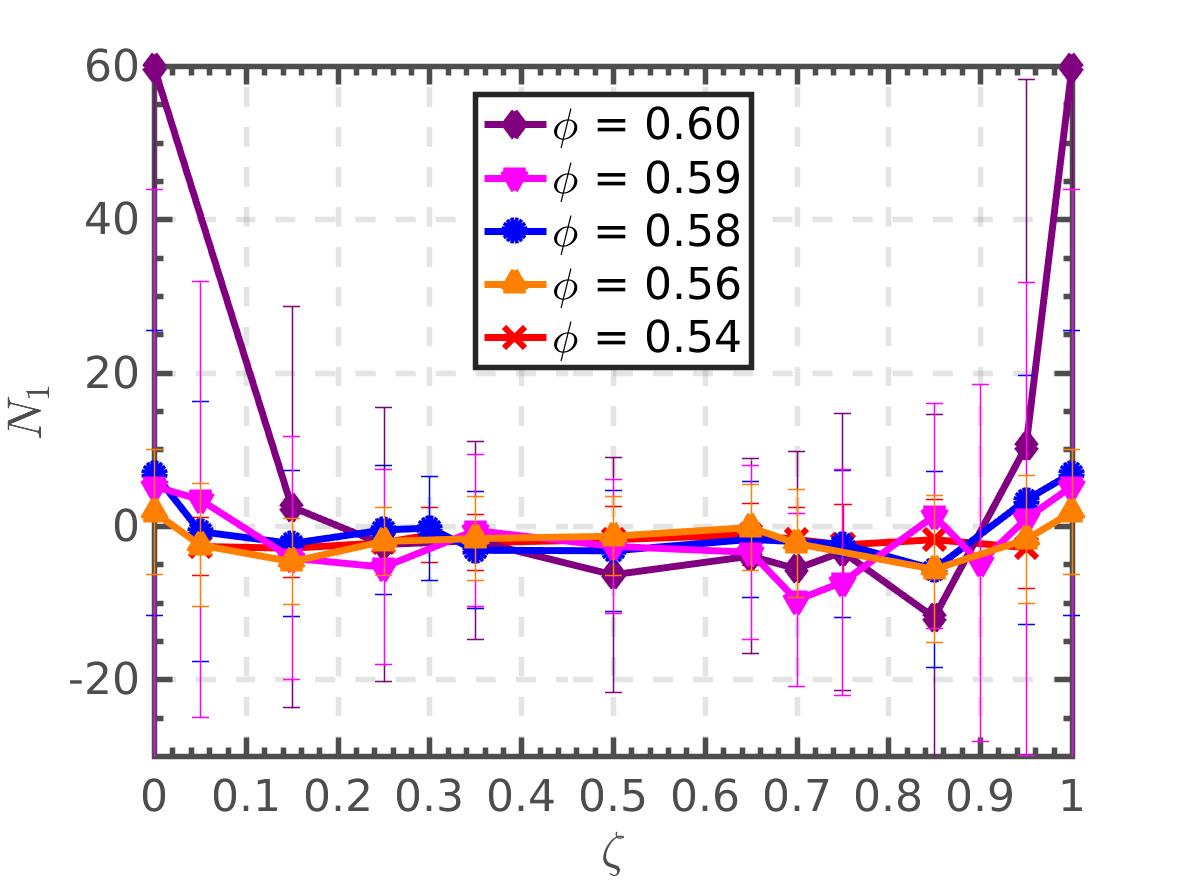} 
    %\caption{Rupture} 
    %\vspace{4ex}
  \end{minipage}\par 
 c) \begin{minipage}[b]{0.42\linewidth}
    \centering
    \includegraphics[width=1.15\linewidth]{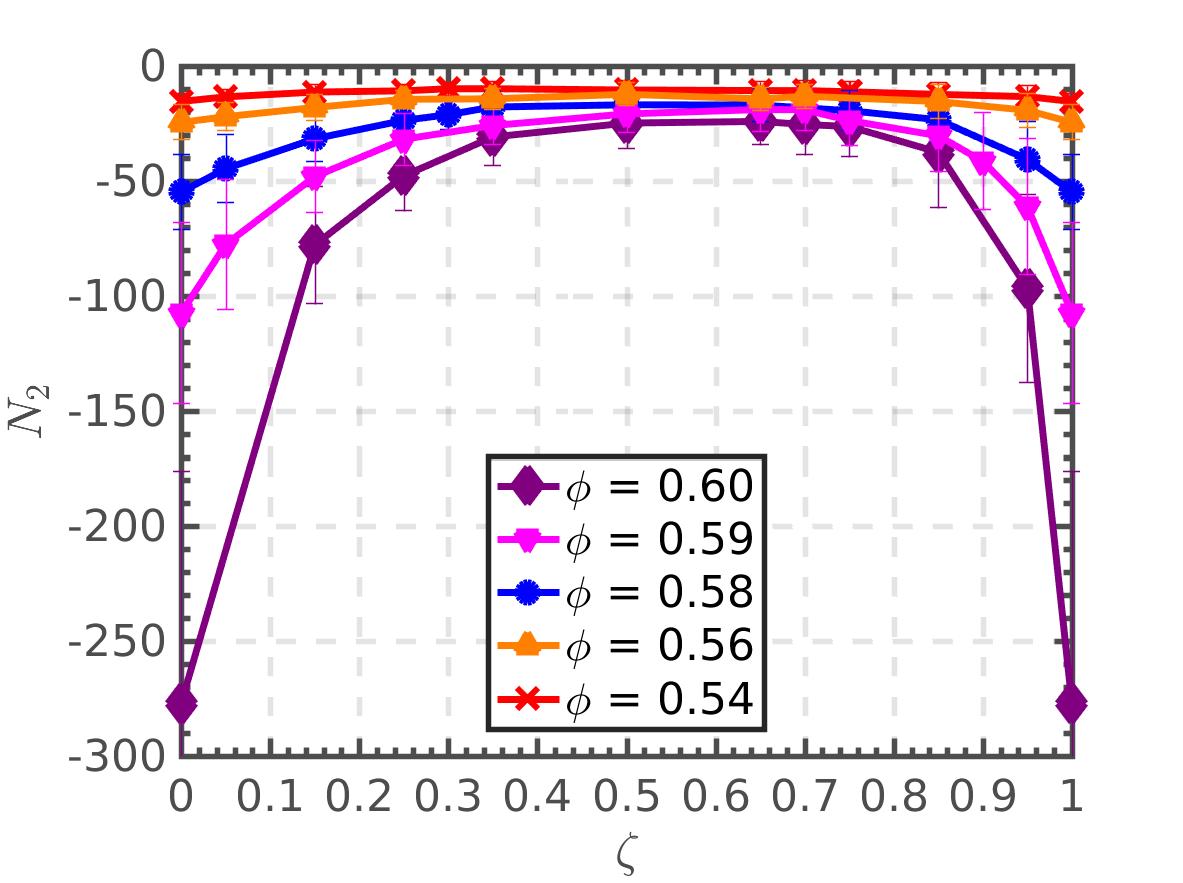} 
    %\caption{DFT, Initial condition} 
   % \vspace{4ex}
  \end{minipage}%% 
  d) \begin{minipage}[b]{0.42\linewidth}
    \centering
    \includegraphics[width=1.15\linewidth]{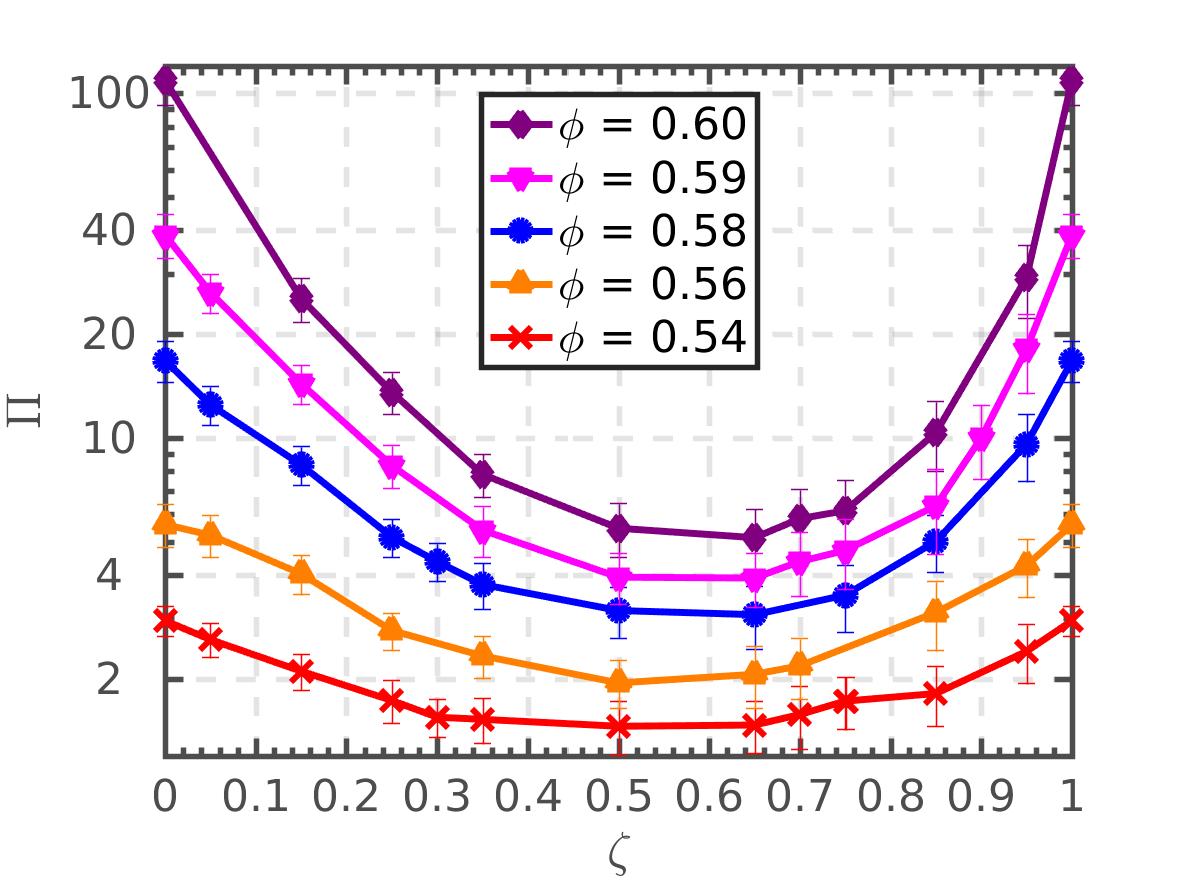} 
    %\caption{DFT, rupture} 
    %\vspace{4ex}
  \end{minipage} 
  \caption{Variation of a) relative viscosity ($\eta_r$), b) $N_1$, c) $N_2$, and d) particle pressure ($\Pi$) with $\zeta$ at different $\phi$ with $\delta = 3$.} 
\label{fgr:bi} 
\end{figure}

\begin{sidewaysfigure}

%\begin{figure*}[!h]
  \centering
%     \subcaptionbox{$\zeta$= 0, 1}[.17\linewidth][c]{%
 %\vspace{-1cm}
 
%\includegraphics[width=.2\linewidth]{Gbb4_1.jpg}}

 %\bigskip
 
     \subcaptionbox{}[.18\linewidth][c]{%
 %\hspace{-1cm}
\includegraphics[width=.18\linewidth]{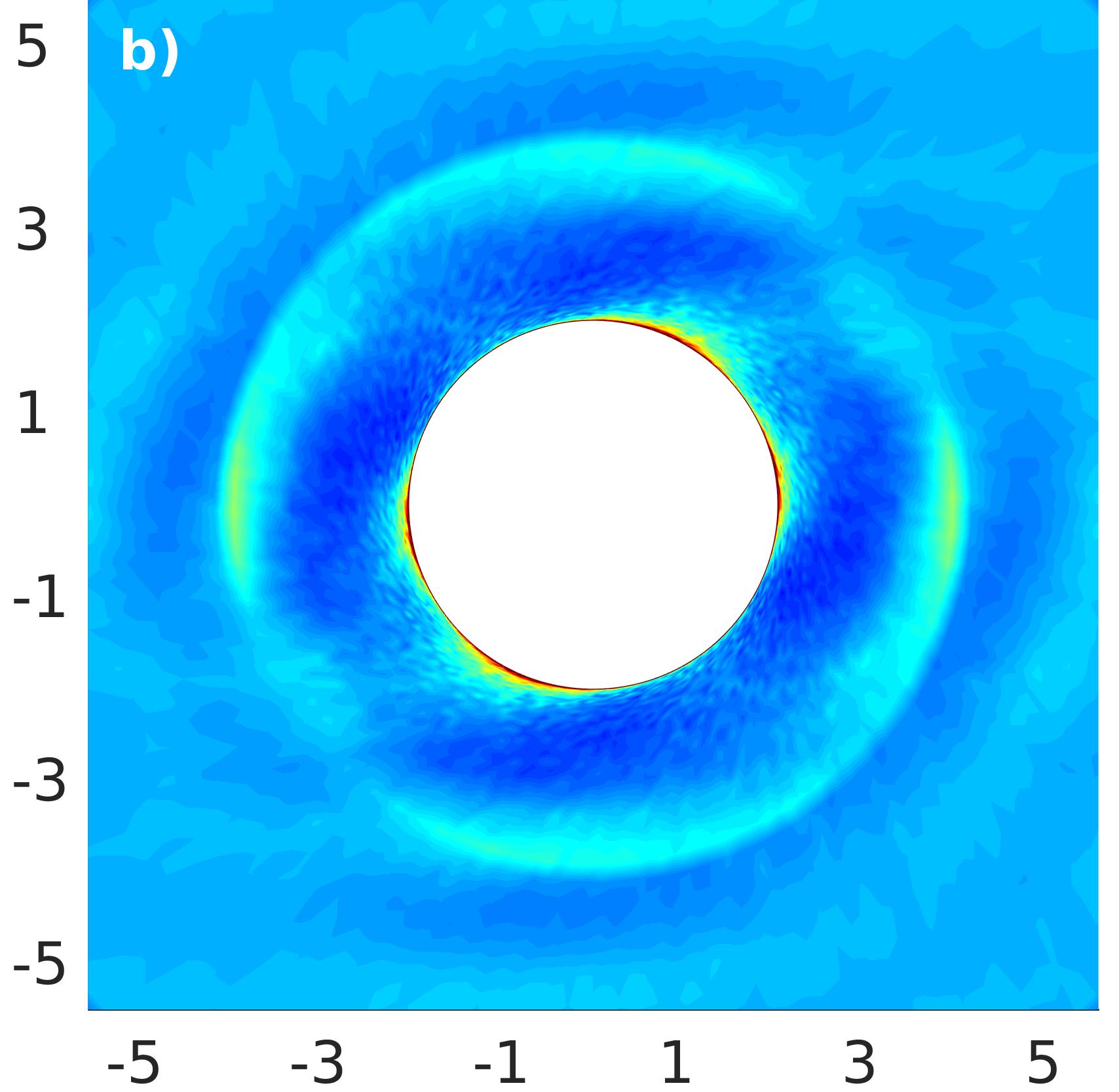}}
  \subcaptionbox{}[0.18\linewidth][c]{%
   
  % \hspace{-2.5cm}   
\includegraphics[width=.18\linewidth]{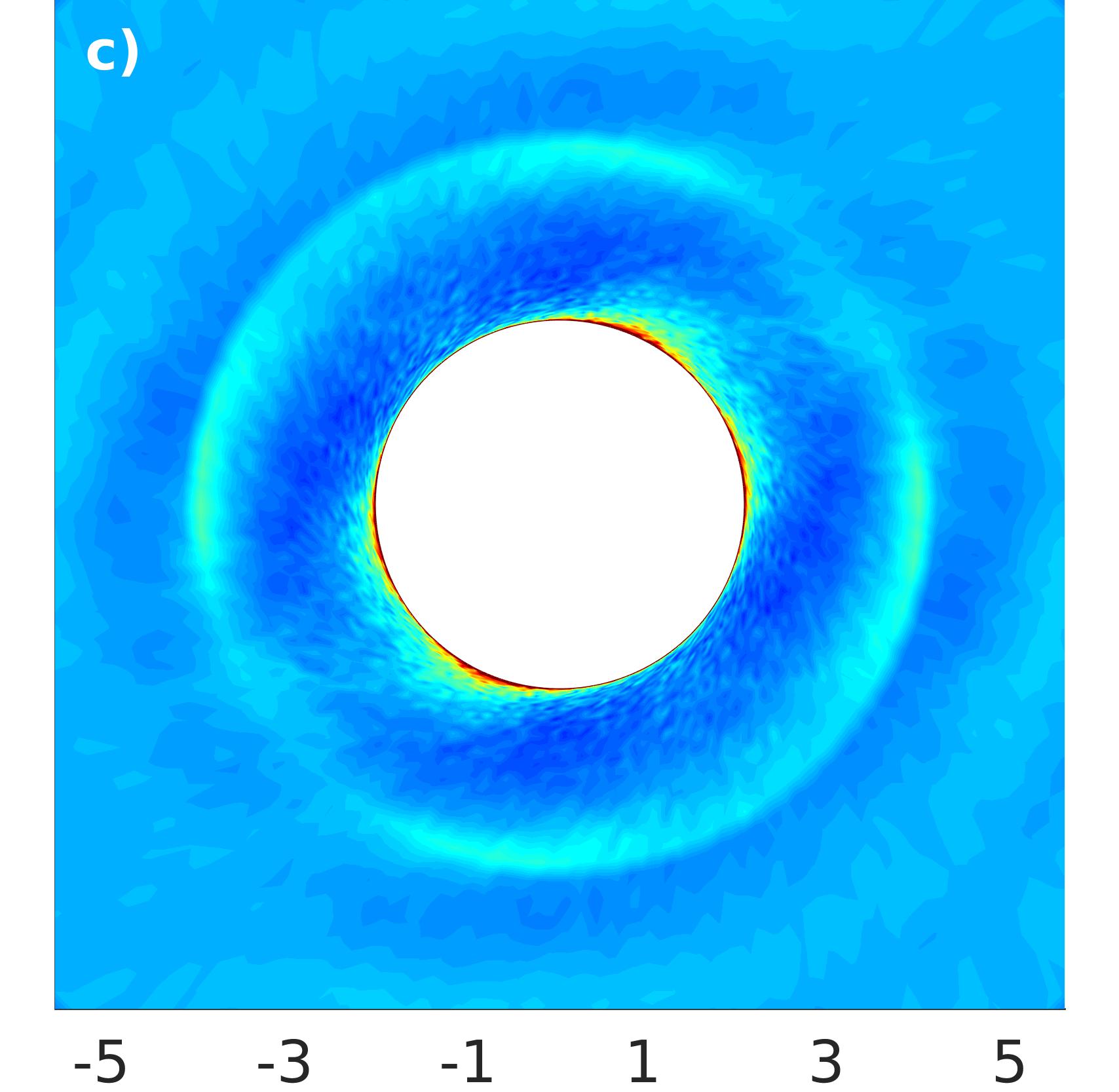}}
  \subcaptionbox{}[.18\linewidth][c]{%
   
%\hspace{-4.5cm}   
\includegraphics[width=.18\linewidth]{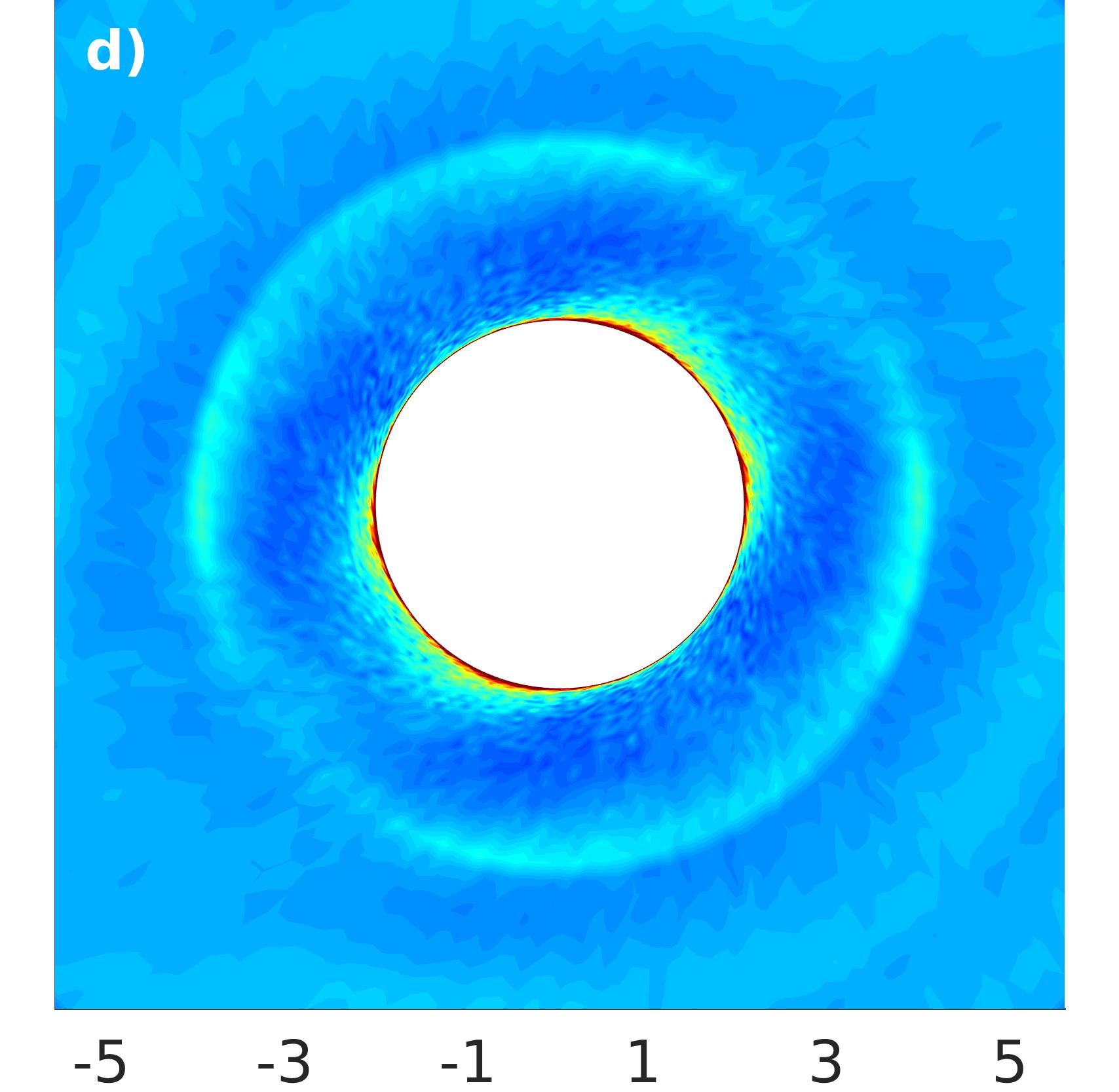}}
\subcaptionbox{}[.18\linewidth][c]{%
   
%\hspace{-5.5cm}   
\includegraphics[width=.18\linewidth]{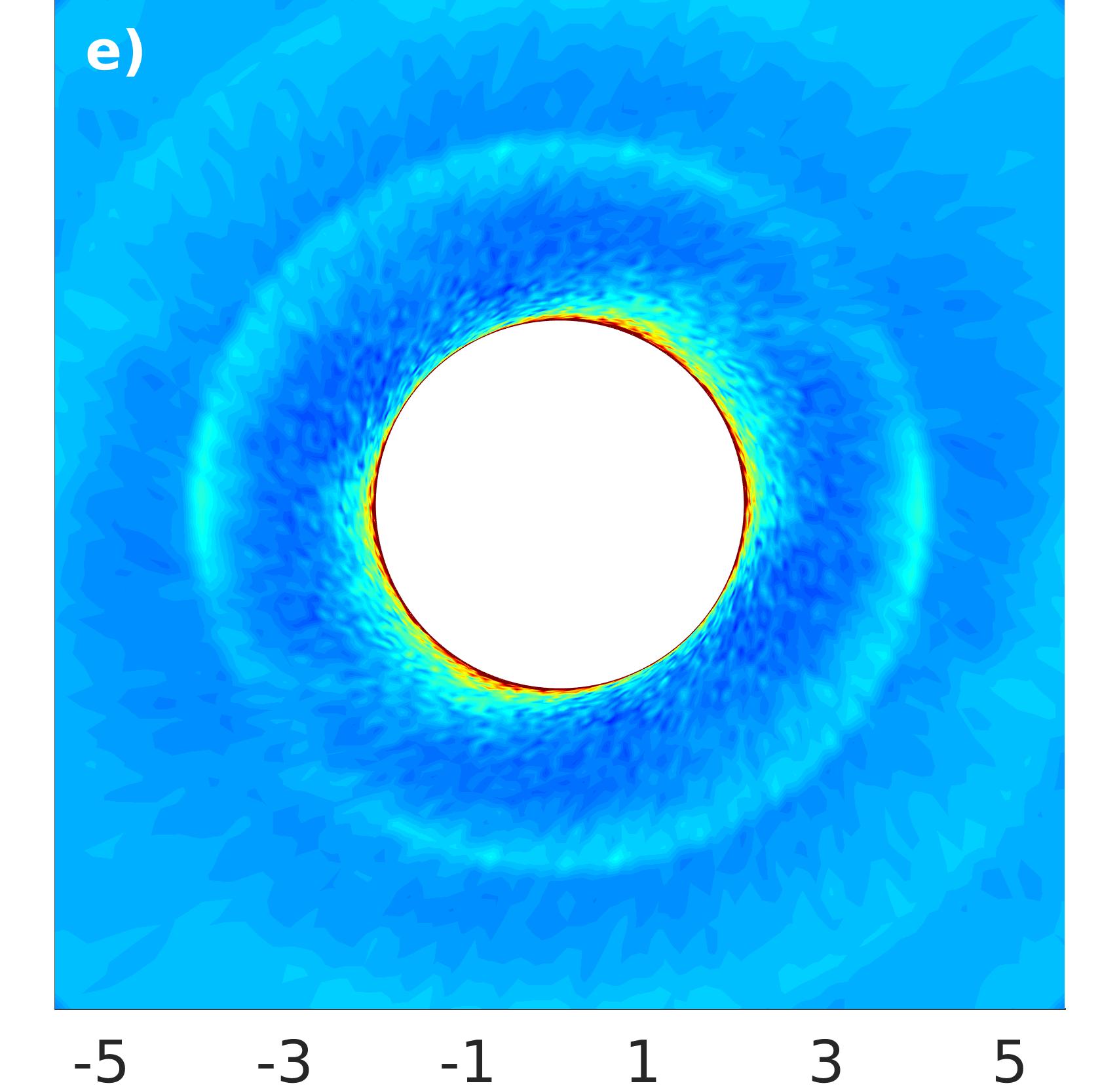}}
  \subcaptionbox{}[.2\linewidth][c]{%
   
%\hspace{-5.5cm}	
\includegraphics[width=.2\linewidth]{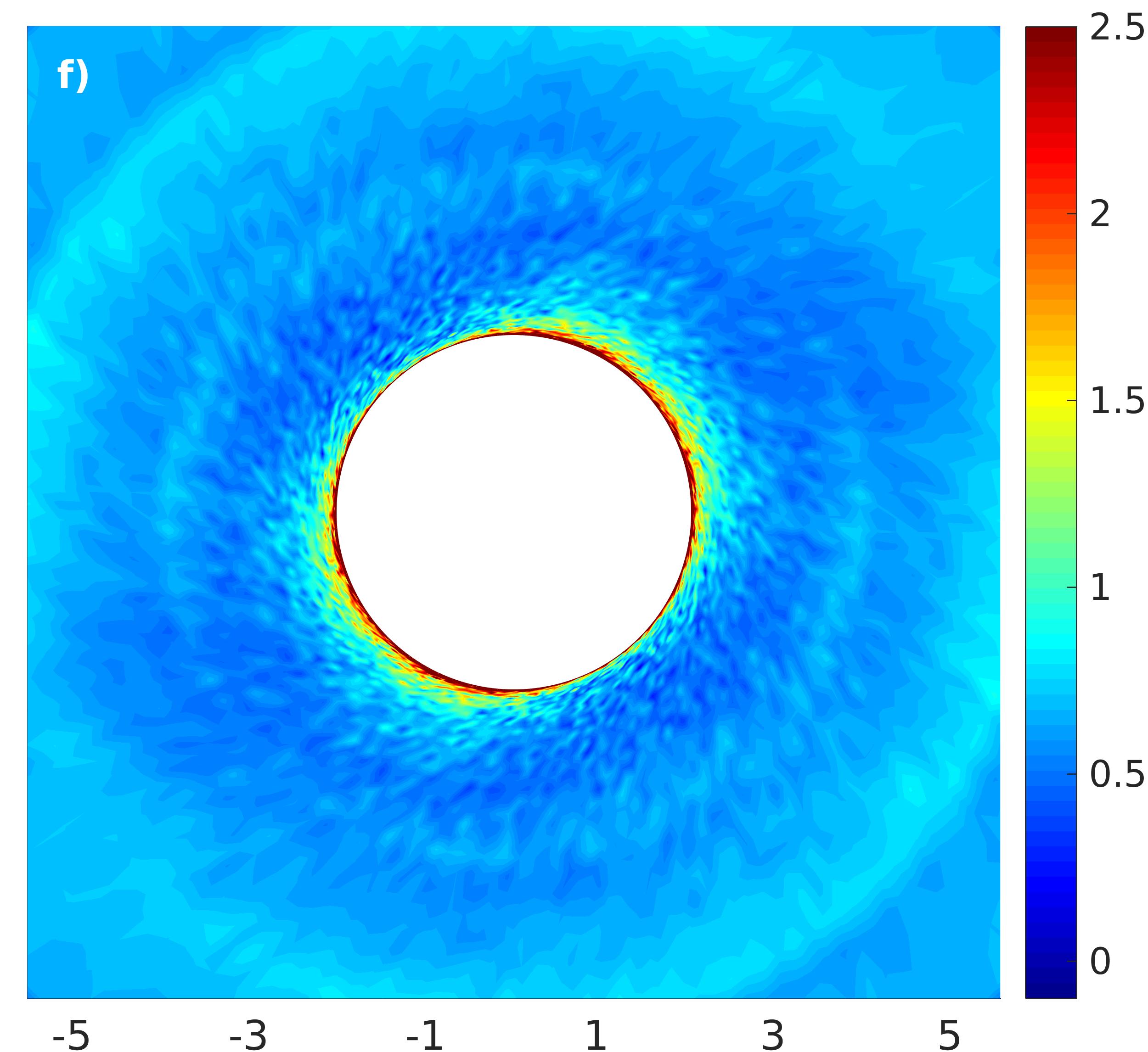}}

  %\bigskip

\subcaptionbox{}[.18\linewidth][c]{%
 %\hspace{-1cm}
  \includegraphics[width=.18\linewidth]{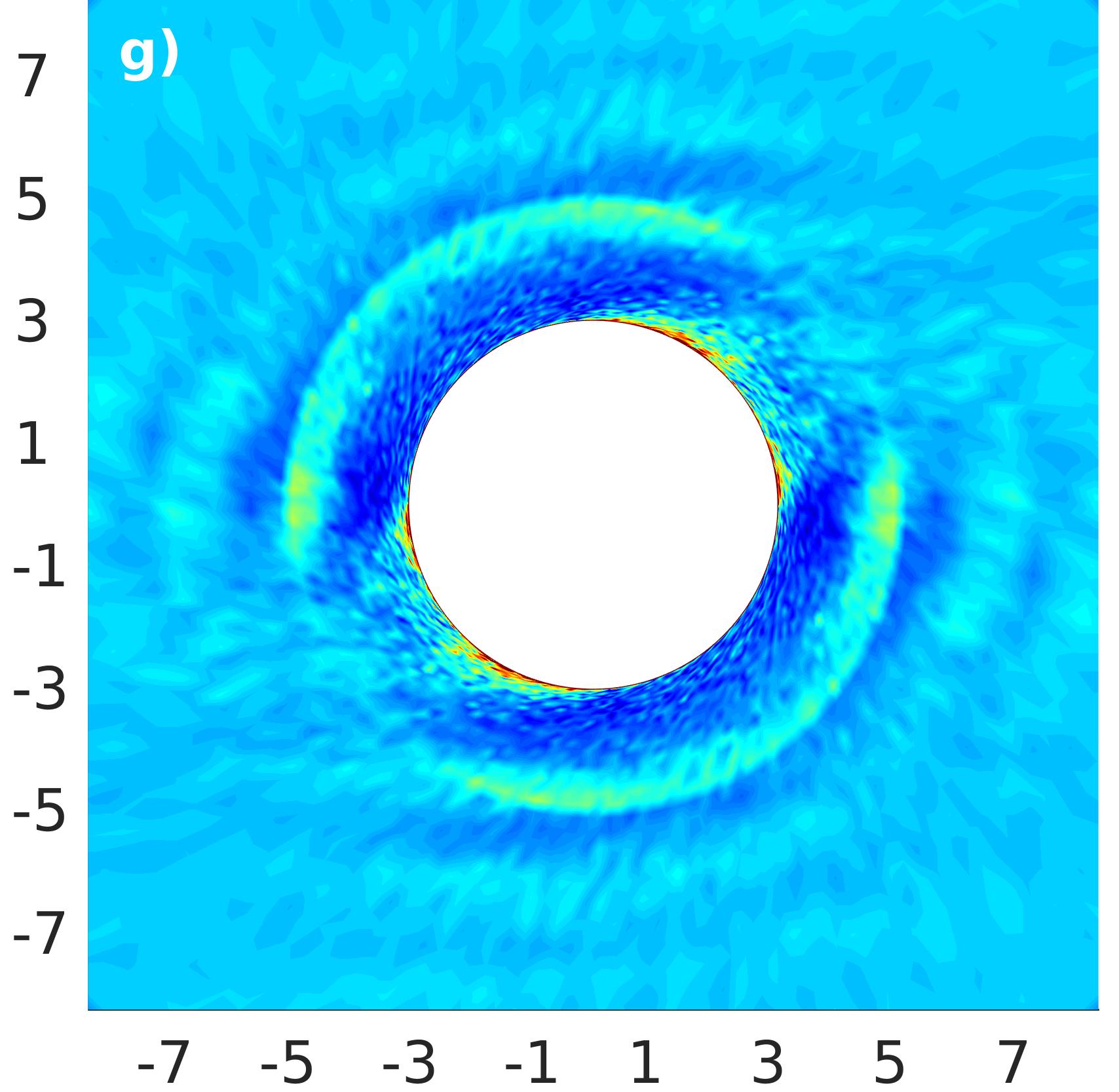}}
  \subcaptionbox{}[.18\linewidth][c]{%
   
   %\hspace{-0.5cm}   
\includegraphics[width=.18\linewidth]{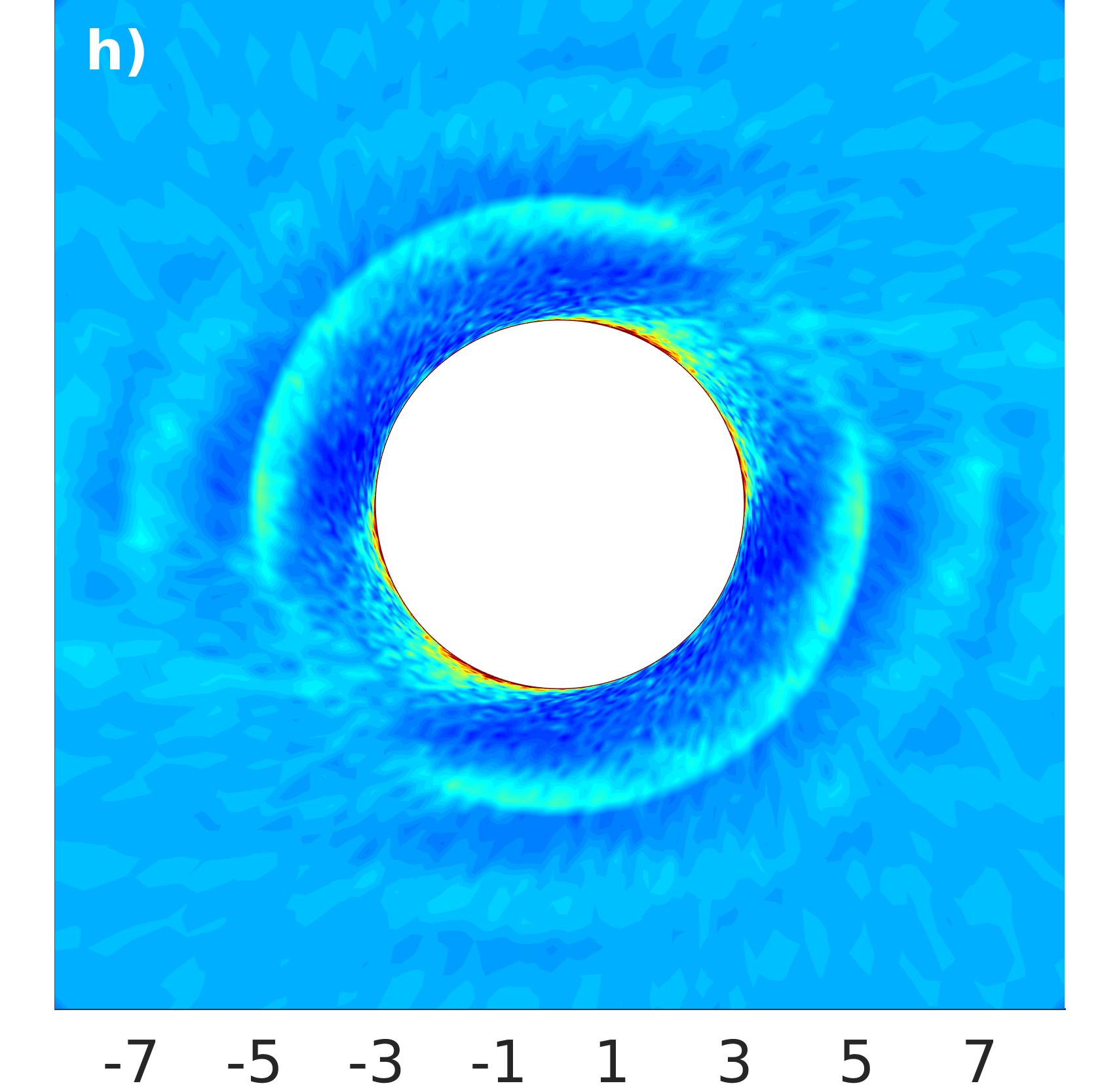}}
  \subcaptionbox{}[.18\linewidth][c]{%
   
%   \hspace{-1cm}   
\includegraphics[width=.18\linewidth]{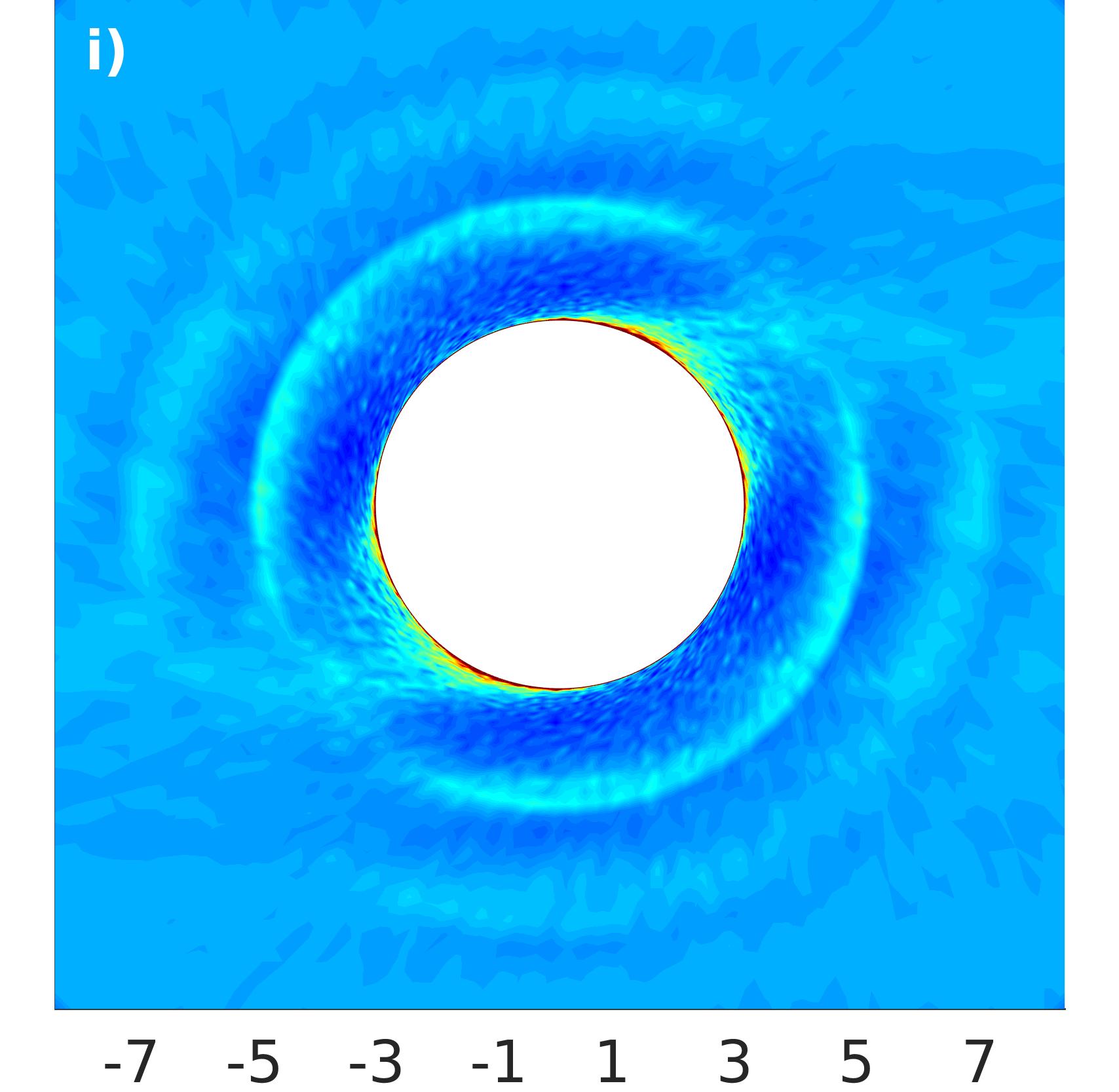}}
\subcaptionbox{}[.18\linewidth][c]{%
   
%\hspace{-1.5cm}   
\includegraphics[width=.18\linewidth]{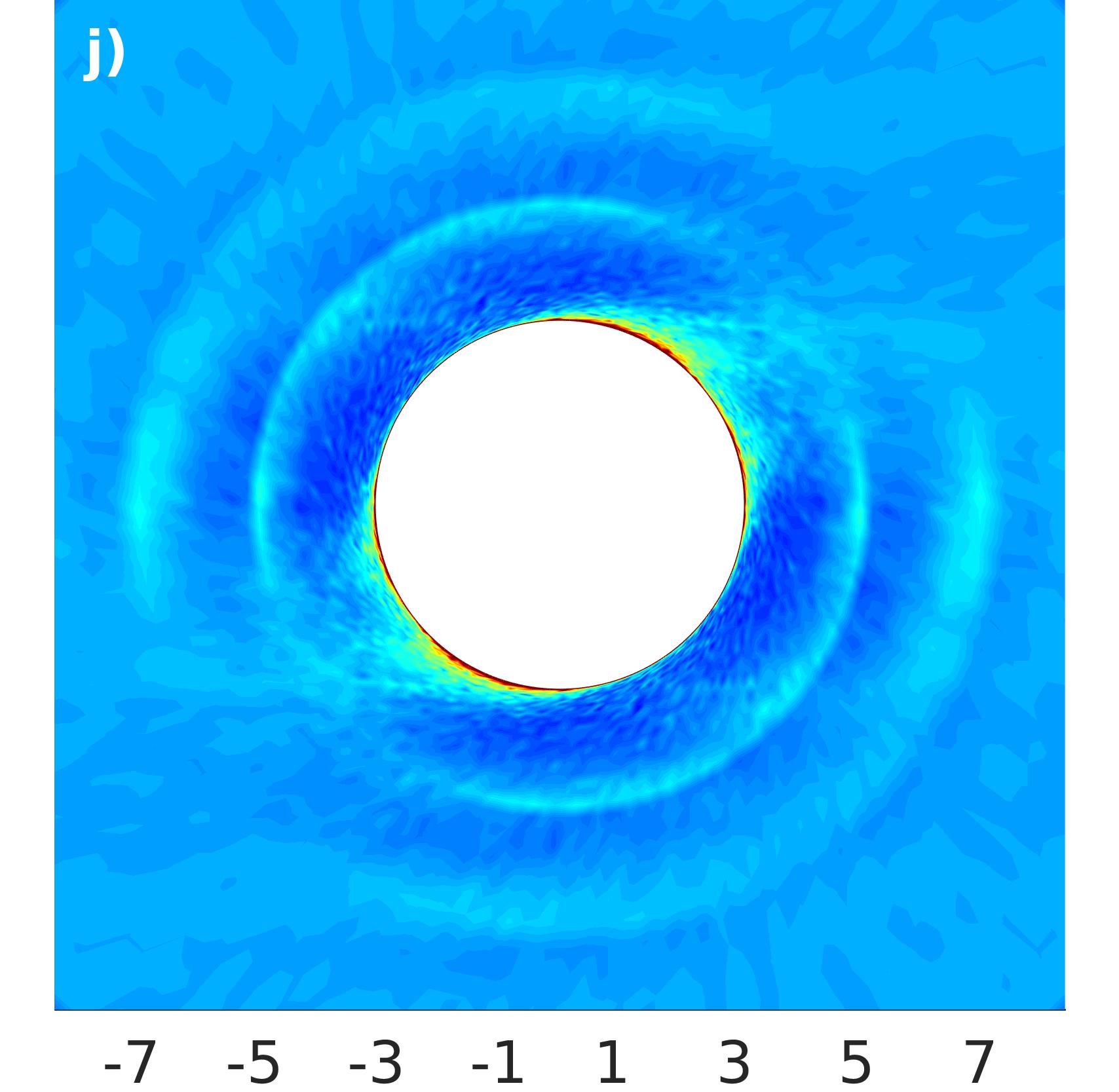}}
  \subcaptionbox{}[.2\linewidth][c]{%
   
%   \hspace{-2cm}   
\includegraphics[width=.2\linewidth]{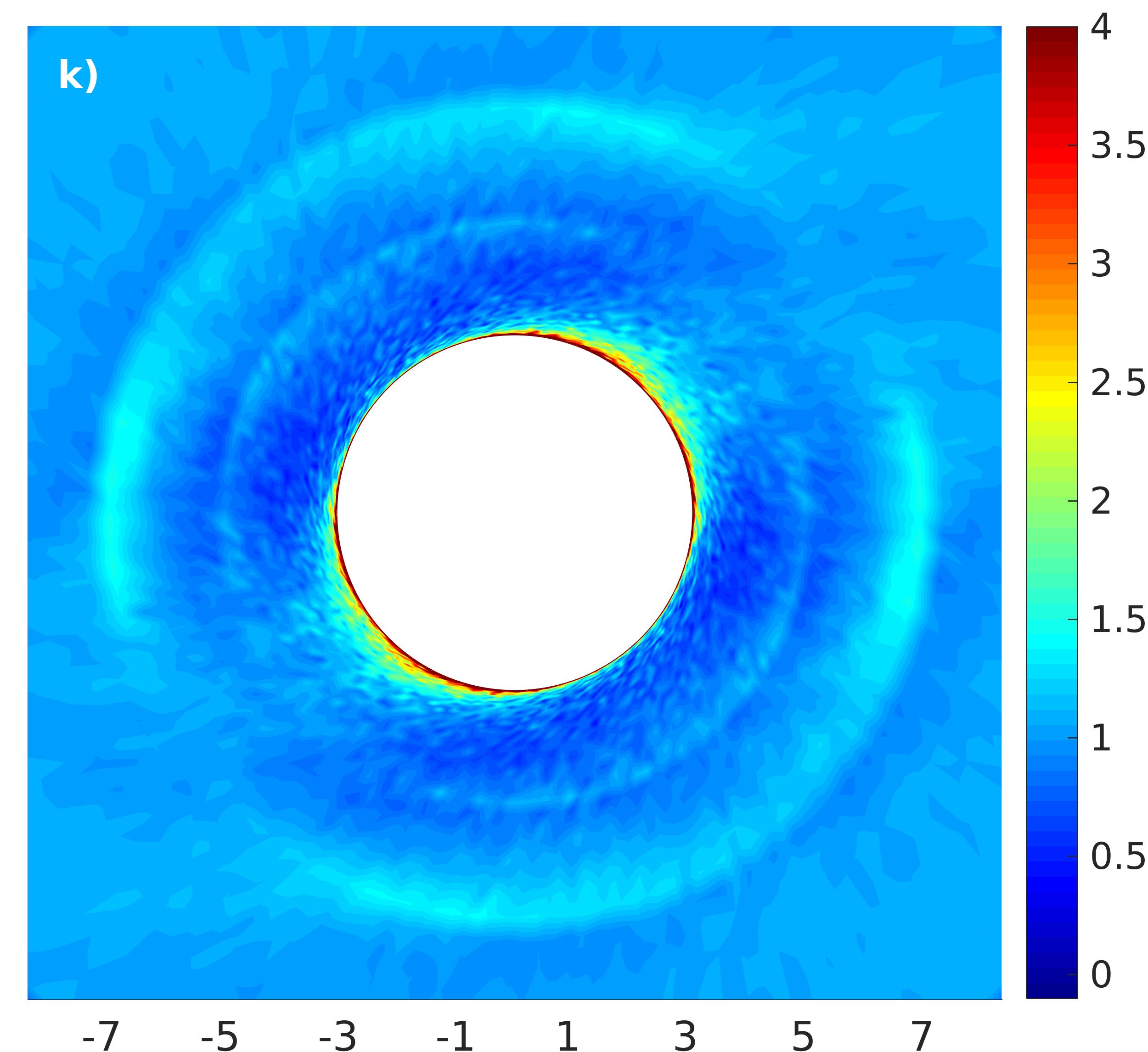}}

  %\bigskip

  \subcaptionbox{$\zeta$= 0.15}[.18\linewidth][c]{%
 %\hspace{-1cm}
\includegraphics[width=.18\linewidth]{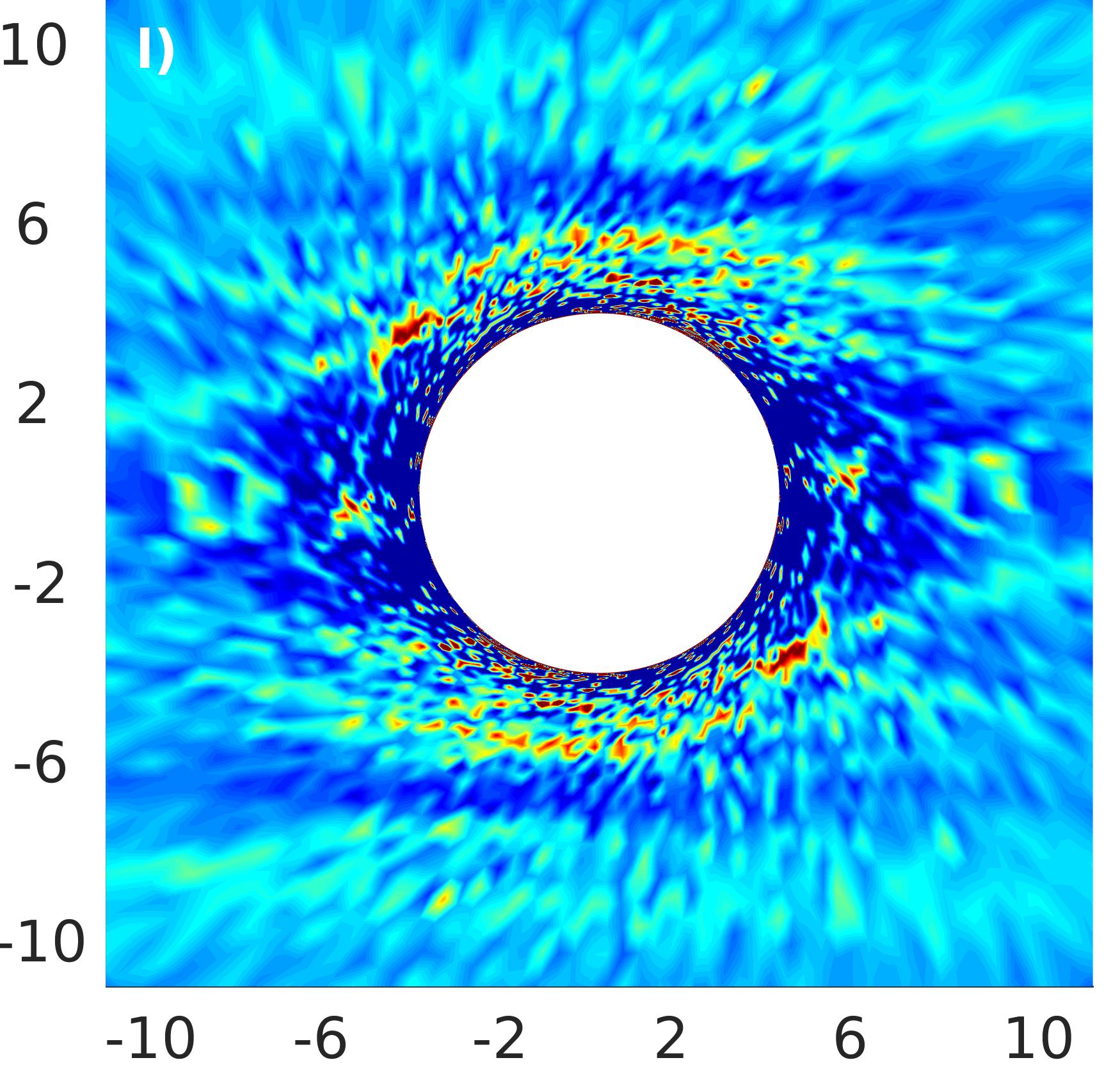}}
  \subcaptionbox{$\zeta$= 0.35}[.18\linewidth][c]{%
   
  % \hspace{-1.5cm}   
\includegraphics[width=.18\linewidth]{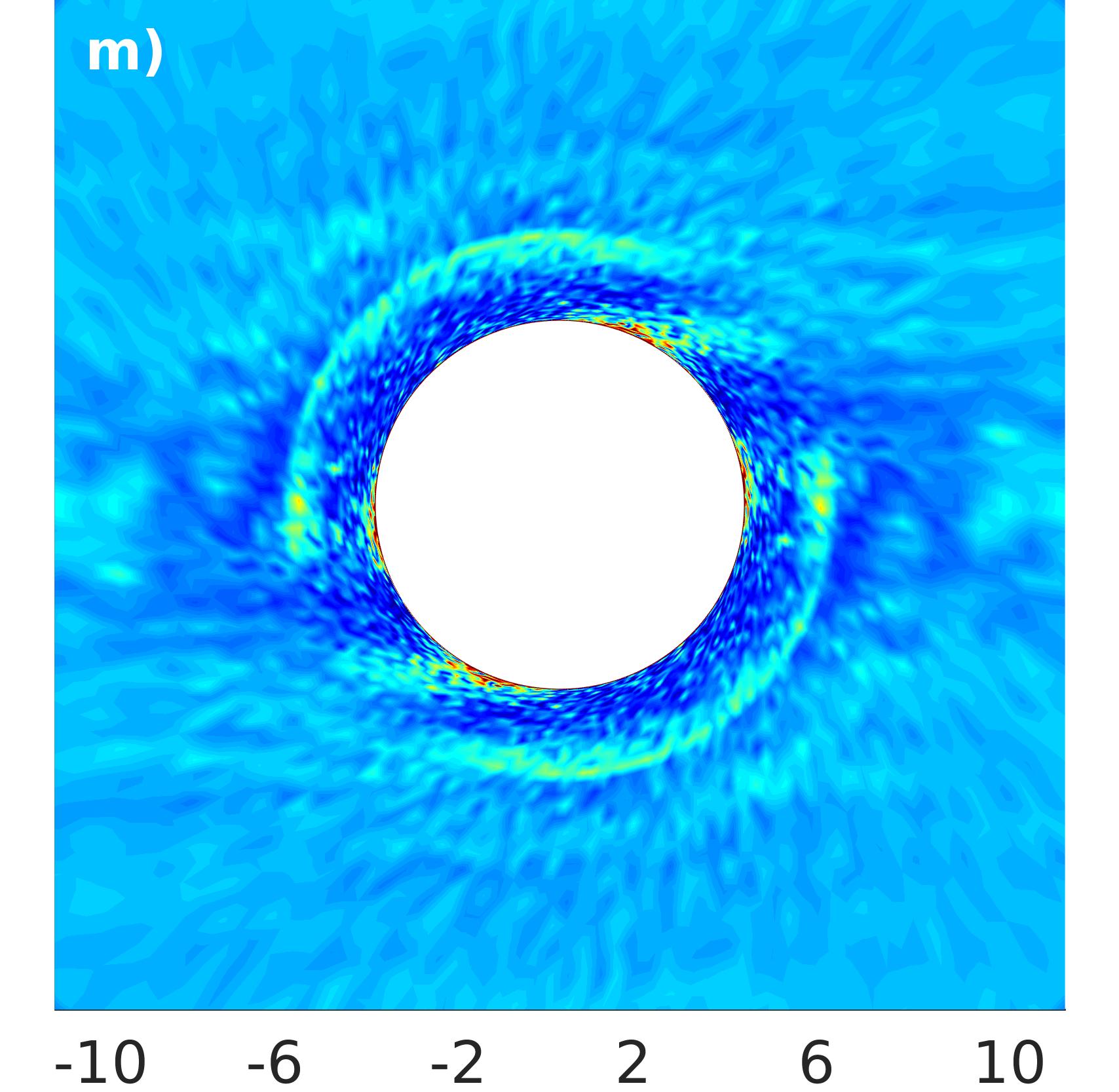}}
  \subcaptionbox{$\zeta$= 0.5}[.18\linewidth][c]{%
   
%   \hspace{-1cm}   
\includegraphics[width=.18\linewidth]{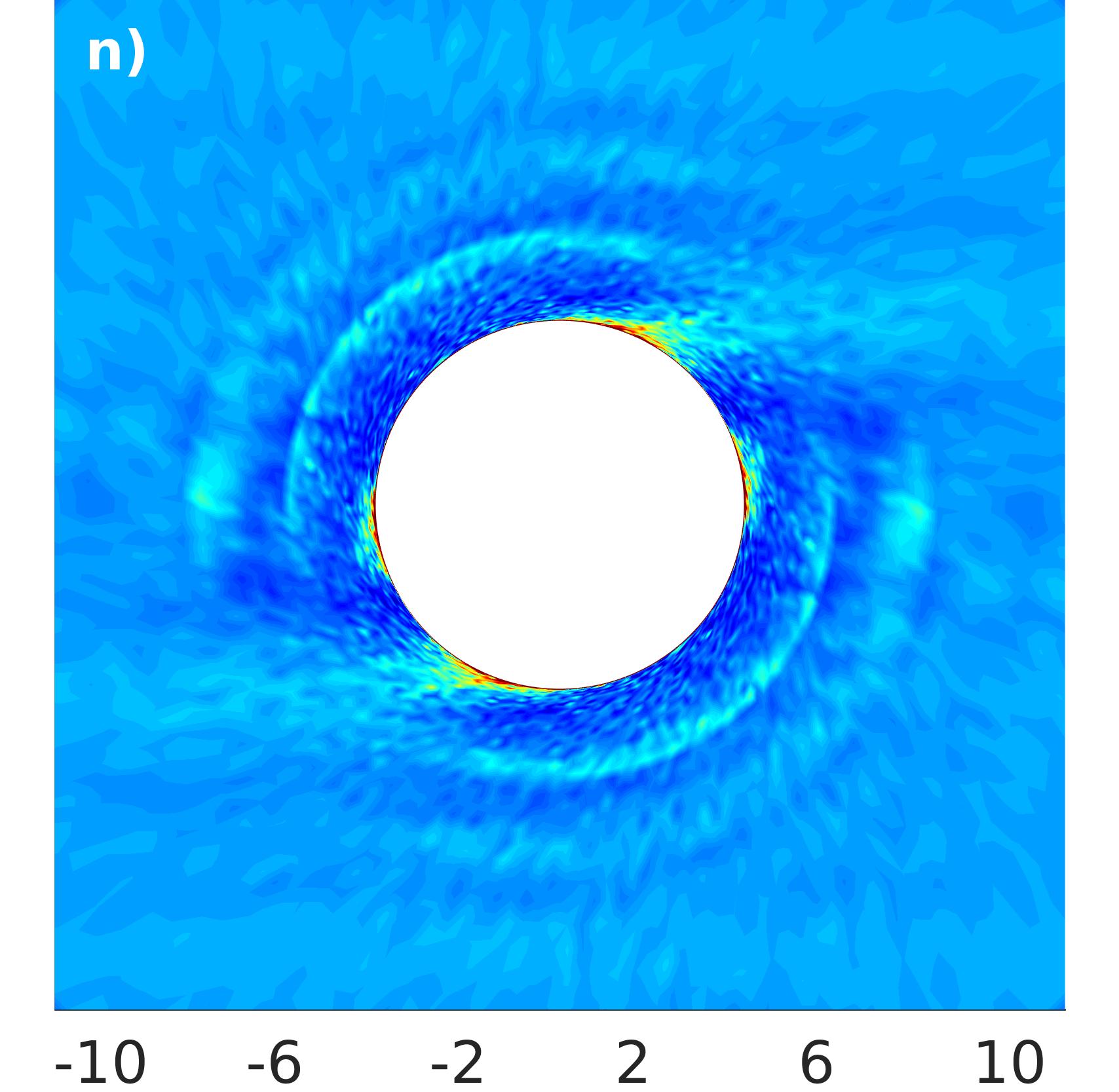}}
\subcaptionbox{$\zeta$= 0.65}[.18\linewidth][c]{%
   
%\hspace{-1.5cm}   
\includegraphics[width=.18\linewidth]{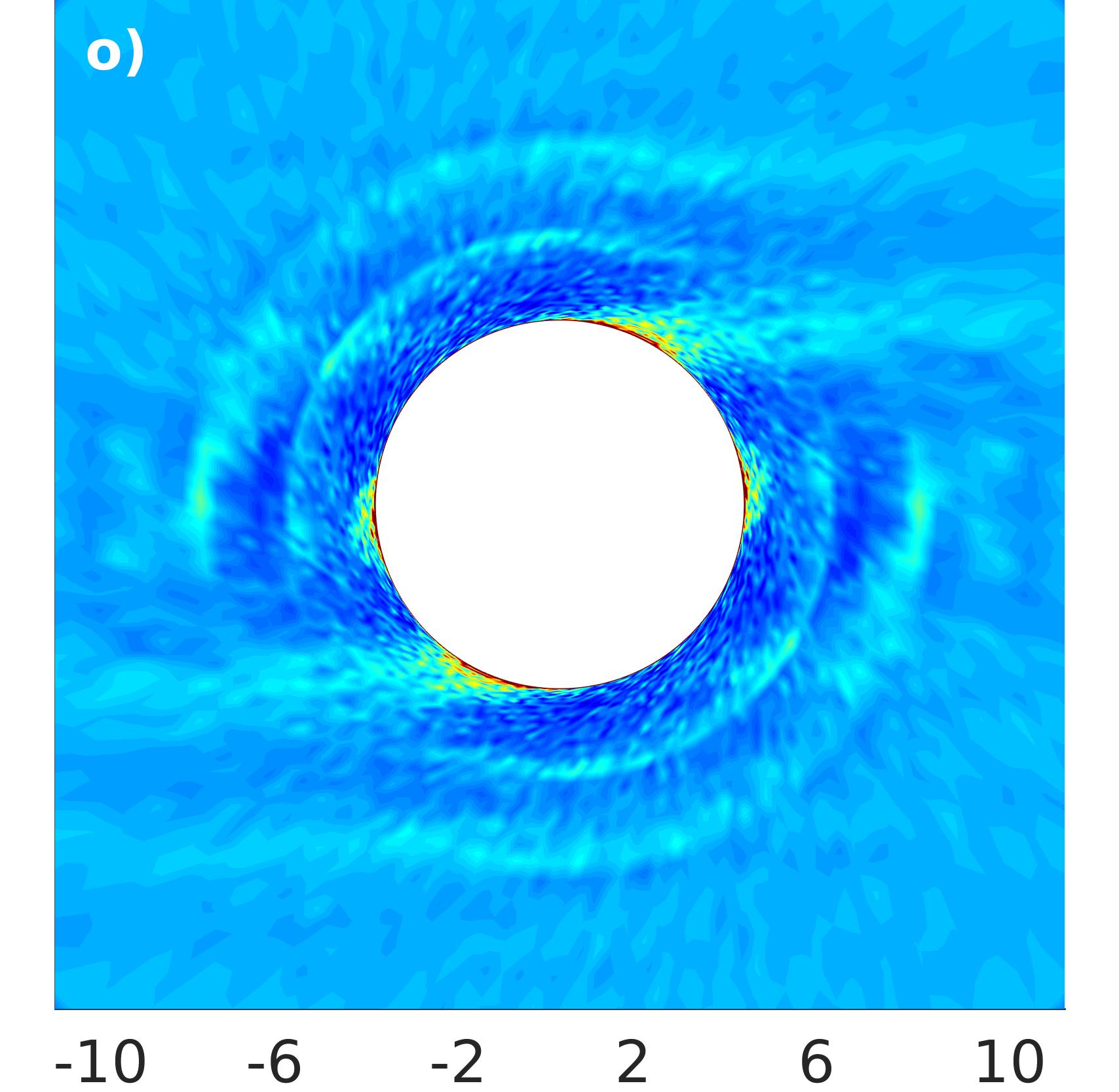}}
  \subcaptionbox{$\zeta$= 0.85}[.2\linewidth][c]{%
   
%   \hspace{-2cm}   
\includegraphics[width=.2\linewidth]{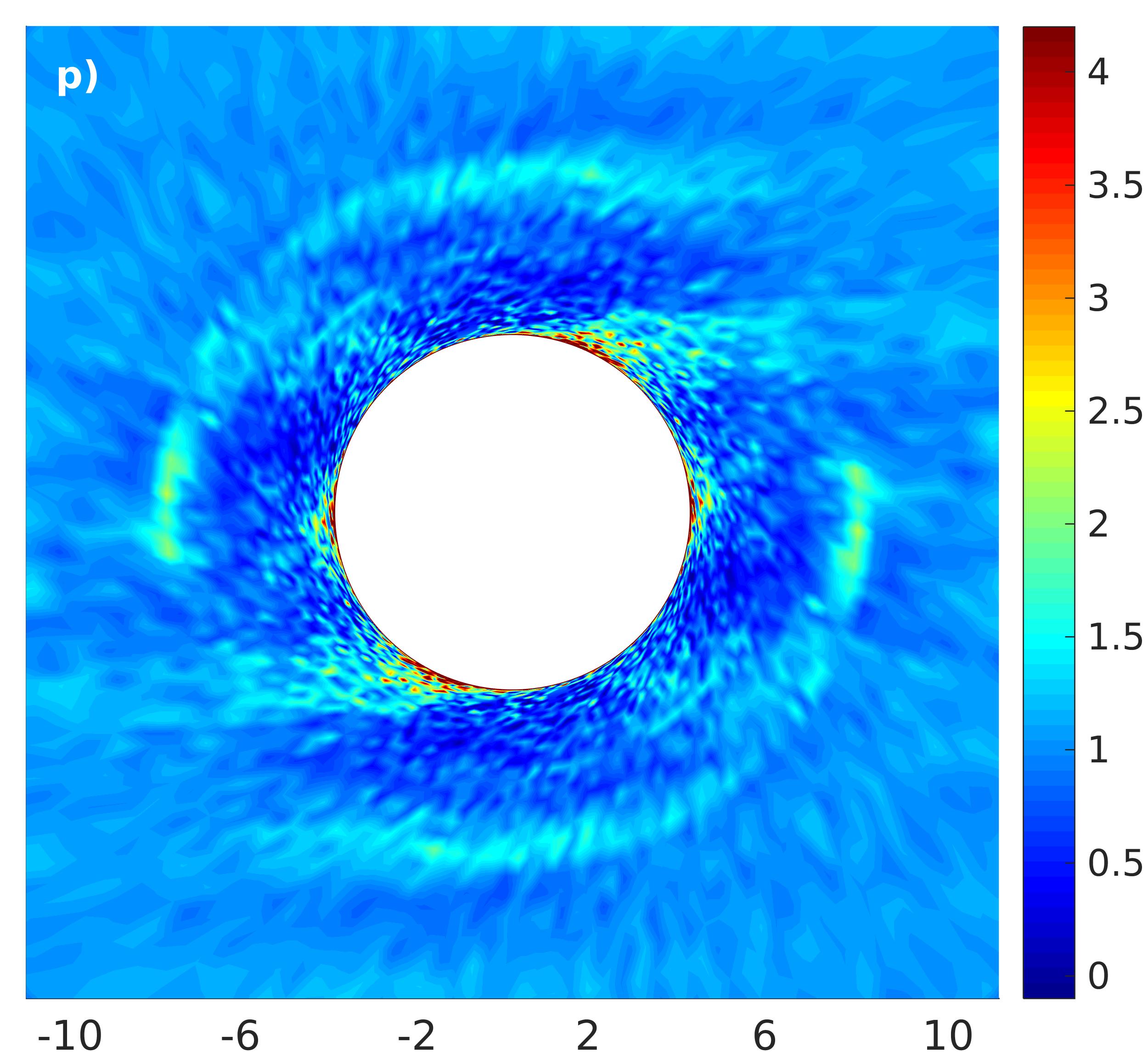}}

  \caption{Pair distribution functions for $\delta$ =2 at $\phi$ = 0.6: row 1 - $g_{s-s}$, row 2 - $g_{l-s}$ and row 3 - $g_{l-l}$ }
  \label{gr}
%\end{figure*}
\end{sidewaysfigure}

\textit{Microstructure}:  We examine the microstructure at various $\zeta$ for a bidisperse suspension.  
 The pair distribution function describes the relative probability of finding a particle at location \textbf{r} away from a reference particle . It is defined here accounting for the probabilities of finding a 
 particle of either size $j$ at a distance from one of size $i$ ($i, j = l$ or $s$):
\begin{equation}
g(\textbf{r})_{i-j} = \frac{P_{j|i}(\textbf{r}|\textbf{0})}{\textit{n}_j}, \label{gr}
\end{equation}

\noindent where $P_{j|i}(\textbf{r}|\textbf{0})$ is the conditional probability of finding a particle $j$ at a distance $\bm{r}$ from a reference particle $i$ at origin and $\textit{n}_j$ is the number density of particle $j$. A recent study [\cite{wang2016spectral}] has examined the pair distribution function of bidisperse suspension at $\delta$ = 2, $\zeta$ = 0.5 for different shear rates (Peclet number) at fixed $\phi$. Here we study bidisperse suspensions ($\delta$ =2) at different $\zeta$ at $\phi =0.6$. Relative distance between particles in the pair distribution is scaled with the small particle radius. Microstructure of monodisperse suspensions is known to demonstrate accumulation of pair probability at contact in the compressional quadrant (2nd and 4th quadrant in the images), and a depletion adjacent to contact in the extensional quadrant (1st and 3rd quadrant in the images), although little data at $\phi = 0.6$ is available. In Fig \ref{gr}b-f we show the small-small particle pair distribution, $g_{s-s}$, at different $\zeta$. In $g_{s-s}$ increased values are pushed into a very narrow `boundary layer' in the compressional quadrant, and elevated but more dispersed probability near contact is seen in the extensional quadrant for both $g_{s-s}$ and $g_{l-s}$. 
The $g_{l-l}$ shows behavior typically seen in monodisperse suspensions, with depletion in the extensional zone. Considering $g_{s-s}$, with increasing $\zeta$, the probability ring corresponding to a small-small-small sequence diminishes and the ring corresponding to a small-large-small sequence is seen to become more prominent. In Fig \ref{gr}g-k we consider the large-large particle distribution function, $g_{l-l}$. Two rings of increased probability beyond contact correspond to the third particle in a large-small-small and large-large-small sequence, respectively. The probability intensity of these rings is observed to change noticeably with $\zeta$. Subtle changes in microstructure are also noticed in $g_{l-l}$ (Fig \ref{gr}l-p) where the distinct probability ring away from contact corresponds to a large-small-large sequence, which decreases with decrease in fraction of small particles.\\

\begin{figure}[!h] 
%\captionsetup{singlelinecheck=false, justification=centering}
  a) \begin{minipage}[b]{0.44\linewidth}
    \centering
    \includegraphics[width=1.15\linewidth]{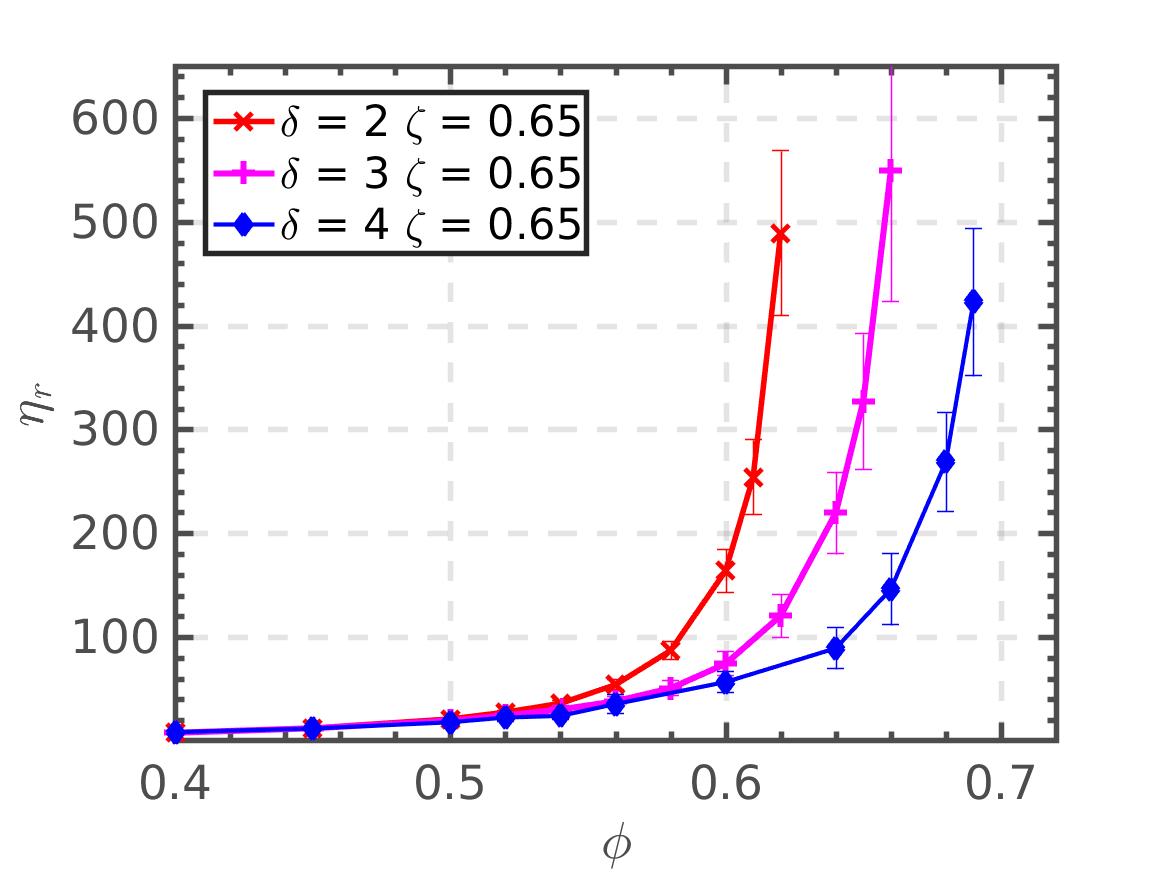} 
    %\caption{Initial condition} 
 %   \vspace{4ex}
  \end{minipage}%%
  b) \begin{minipage}[b]{0.44\linewidth}
    \centering
    \includegraphics[width=1.15\linewidth]{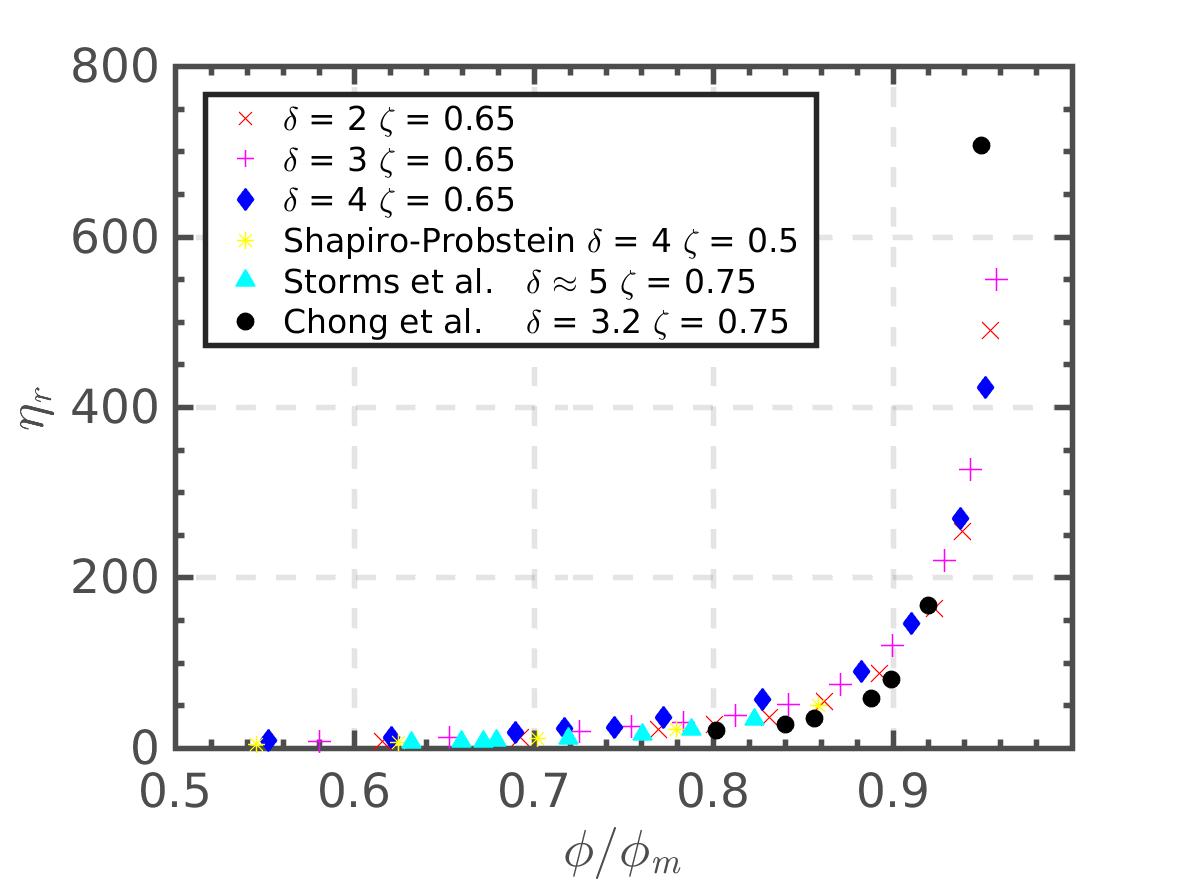} 
    %\caption{Rupture} 
%    \vspace{4ex}
  \end{minipage}\par 
  \caption{a) Relative viscosity as a function of $\phi$ for different bidisperse suspensions.  b) Viscosity curves replotted against reduced volume fraction $\phi/\phi_m$, with $\phi_m$ values estimated by fitting curves with  (\ref{one}). } 
\label{fgr:collapse} 
  \end{figure}

\textit{Relative viscosity collapse}: In Fig. \ref{fgr:collapse}a we plot the relative viscosity ($\eta_r$) against $\phi$ for suspensions of $\delta = 2$, 3, and 4, all at $\zeta =0.65$, which is near $\zeta_{min}$ (the composition with lowest relative viscosity). Note that we do not consider viscosity for solid fractions below $\phi = 0.4$, because the simulation algorithm does not account for far-field hydrodynamic interactions and these are expected to have 
appreciable influence at $\phi <0.4$. The figure demonstrates an increase in maximum packing (or jamming point), $\phi_m$, with increasing $\delta$: the singularity of the viscosity function is pushed towards higher $\phi$.  These results show how the increase in $\phi_m$ corresponds to a decrease in viscosity at fixed $\phi$. Next, we fit the viscosity versus $\phi$ curves of Fig. \ref{fgr:collapse}a to Eq. \ref{one} to determine $\phi_m$. In Fig. \ref{fgr:collapse}b, $\eta_r$ is replotted against the reduced volume fraction, $\phi/\phi_m$ for the data of Fig. \ref{fgr:collapse}a, largely collapsing the curves for different size ratios.

\subsection{\label{sec:level2}Polydisperse suspensions}

\subsubsection{\label{sec:level3}Parameterization}

Here we study the rheology of polydisperse suspensions whose linear size (i.e. radius) number distributions follow normal or log-normal forms.  The mean size of the particles in the polydisperse distributions that are simulated is defined as the reference size, and thus in scaled form is set to $\langle$$a$$\rangle =1$. These suspensions are characterized by their polydispersity $\alpha$, given by (\ref{polyd}). As discussed above, the maximum $\alpha$ is limited by our choice of largest ($a_{max}$) and smallest ($a_{min}$) particle sizes to maintain $a_{max}/a_{min} \le 4.0$, and hence the maximum is $\alpha = 0.2$ and $\alpha$ = 0.3 for normal and log-normal distributions, respectively. \\

\begin{figure}[!h] 
%\captionsetup{singlelinecheck=false, justification=centering}
  a) \begin{minipage}[b]{0.44\linewidth}
    \centering
    \includegraphics[width=1.1\linewidth]{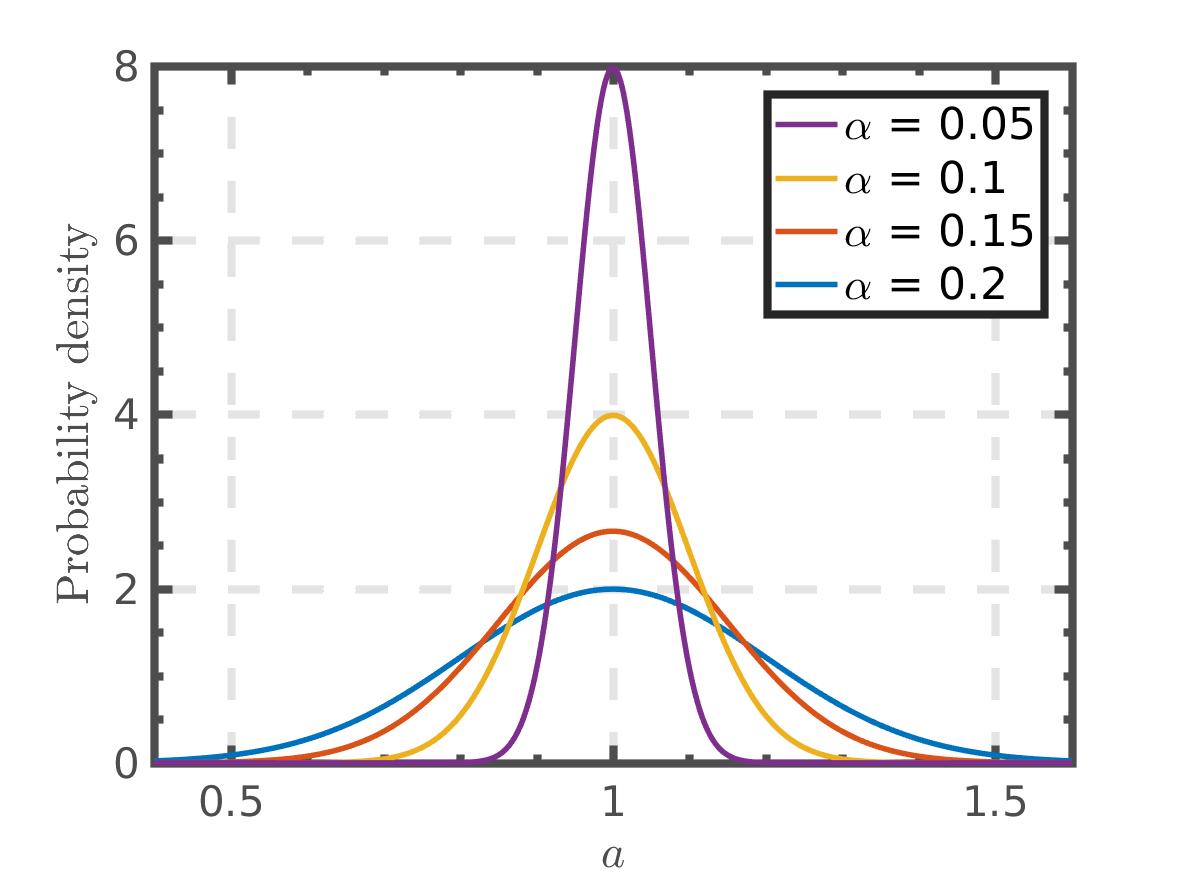} 
    %\caption{Initial condition} 
    %\vspace{4ex}
  \end{minipage}%%
  b) \begin{minipage}[b]{0.44\linewidth}
    \centering
    \includegraphics[width=1.1\linewidth]{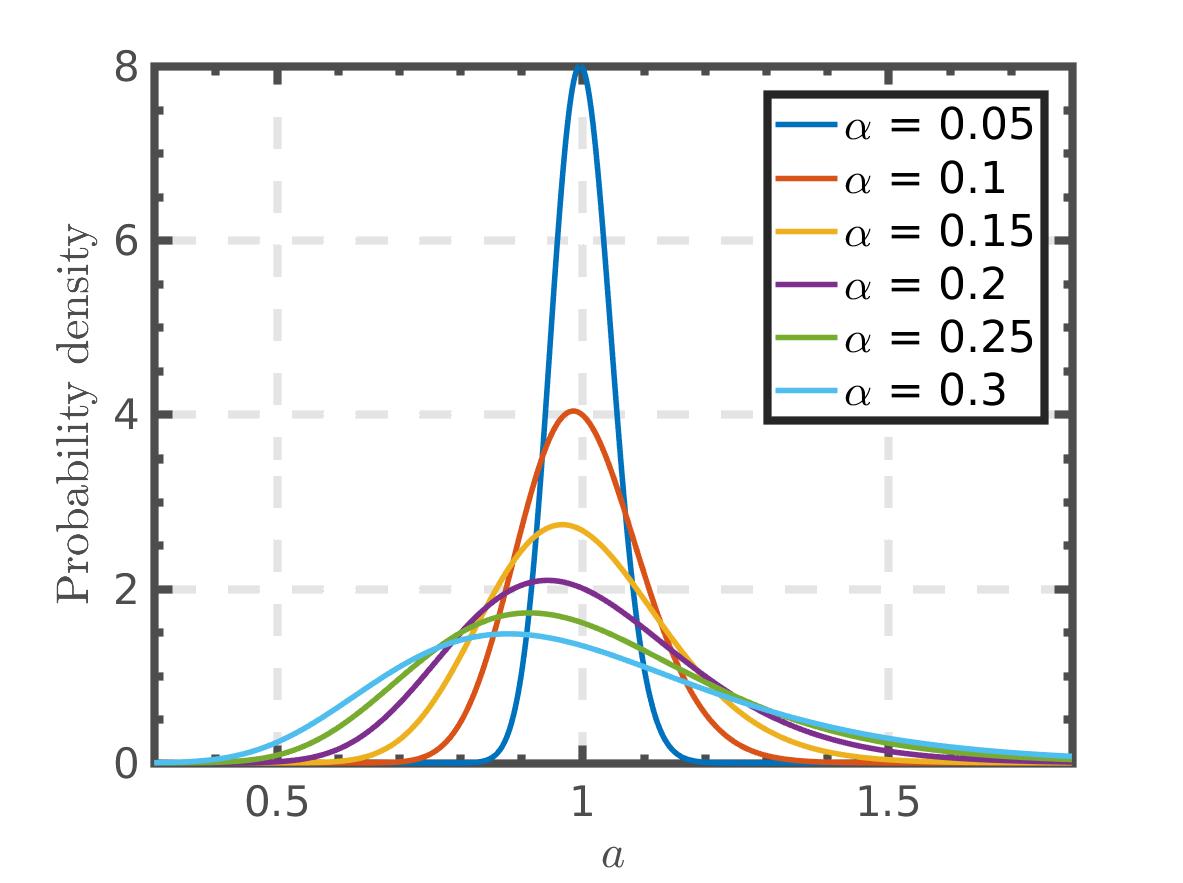} 
    %\caption{Rupture} 
   % \vspace{4ex}
  \end{minipage}\par 
  \caption{Probability density functions for a) normal and b) log-normal distributions for different polydispersity ($\alpha$)}.
\label{fgr:polypdf} 
  \end{figure}

\begin{table}[b]
\caption{\label{table1}Statistical parameters of normal and log-normal distributions distributed according to their radii.}
\begin{ruledtabular}
\begin{tabular}{lcr}
%&\multicolumn{2}{c}{Normal}\\
&Normal distribution&Log-normal distribution\\
\hline
Probability density function & $\frac{1}{\sigma \sqrt{2\pi}}e^{-(x-\mu)^2/2\sigma^2}$ & $\frac{1}{x \sigma \sqrt{2\pi}}e^{-(\ln{x}-\mu)^2/{2\sigma^2}}$\\
Mean & $\mu$ & $e^{(\mu + \sigma^2/2)}$\\
Polydispersity ($\alpha$) & $\sigma$ & $\sqrt{(e^{\sigma^2}-1)e^{2\mu + \sigma^2}}$\\
Skewness ($S$) & 0 & $(e^{\sigma^2}+2)\sqrt{e^{\sigma^2}-1}$\\
\end{tabular}
\end{ruledtabular}
\end{table}

\noindent The skewness factor ($S$), which is the third standardized moment of a distribution, is also used to characterize the polydisperse suspensions. Defined by 
\begin{equation}
%S = \langle \Delta R^3 \rangle/\langle \Delta R^2 \rangle^{3/2}, \label{4}
S = \langle \Delta a^3 \rangle/\langle \Delta a^2 \rangle^{3/2}, \label{4}
\end{equation}
for particles with variable radius $a$, the skewness gives a measure of the asymmetry around the mean. Particles having a normal distribution have $S = 0$, while log-normal distributions have non-zero skewness.\\

By varying the polydispersity factor $\alpha$, we systematically study the rheology of different normal and log-normal suspensions, the latter of which also have an associated skewness ($S$). The probability density functions for the polydisperse suspensions studied are plotted in Fig. \ref{fgr:polypdf}a-b. The mathematical expressions for the relevant statistical parameters (mean, polydispersity and skewness) of the distributions are tabulated in Table \ref{table1}. Here $\mu$ and $\sigma$ take on their usual statistical meaning of the mean and standard deviation in a normal distribution, respectively.
Note that the kurtosis and other higher moment statistical descriptions were found in other work to not be influential on the maximum packing [\cite{desmond2014influence}] and hence are not taken into consideration. \\

\begin{figure}[t]
\centering
\includegraphics[width=.5\linewidth]{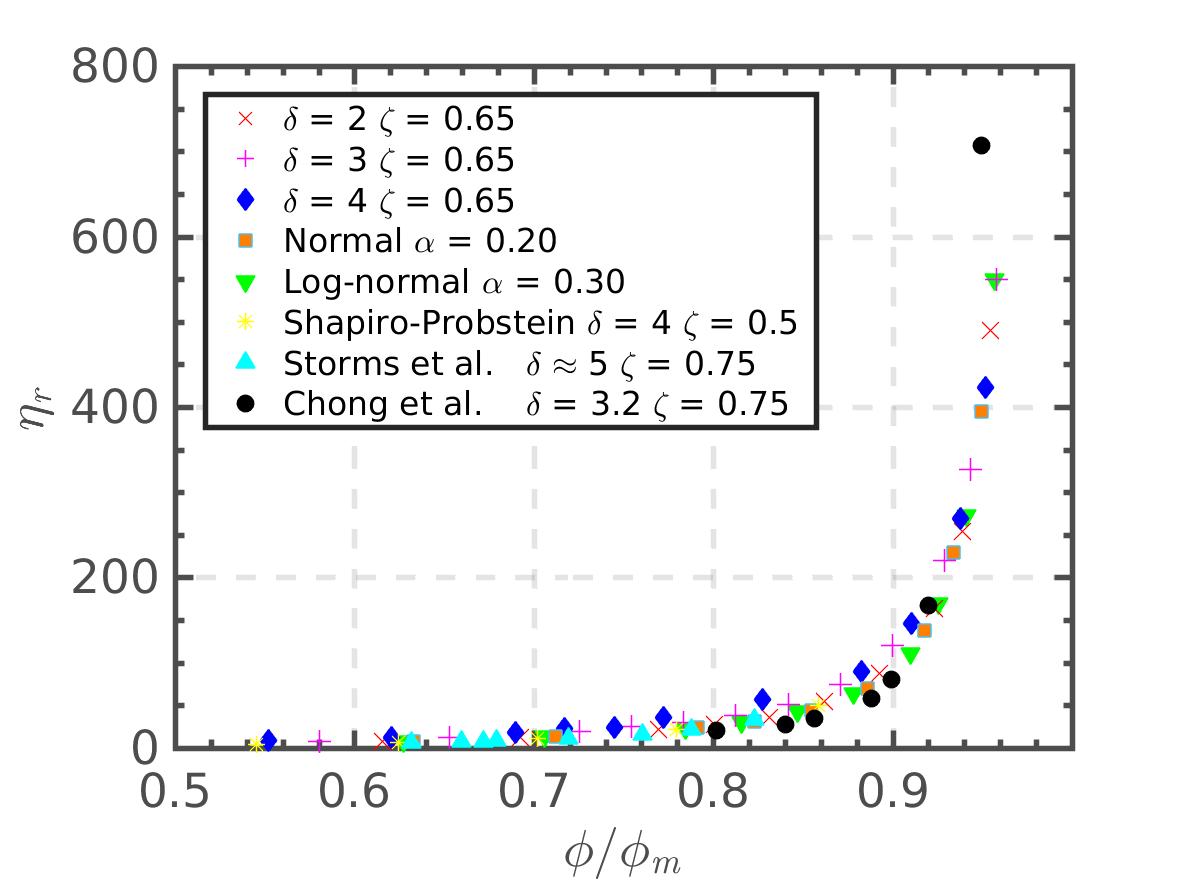}
\caption{The viscosity curves from Fig. \ref{fgr:collapse}b replotted with normal and log-normal suspensions.}
\label{fig:collapse2}
\end{figure}

The $\mu$ and $\sigma$ values for normal and log-normal systems are calculated from the equations in Table \ref{table1} by simultaneously solving the respective expressions of mean and required polydispersity ($\alpha$). Polydisperse distributions are then generated using the calculated $\mu$ and $\sigma$ as described in Sec. \ref{model}. The associated skewness for these systems is calculated by substituting $\mu$ and $\sigma$ in the expressions in Table \ref{table1}. We use this skewness to generate statistically equivalent bidisperse systems having similar maximum packings. Relative viscosity curves of log-normal and normal suspensions are also seen to collapse with reduced volume fraction in Fig. \ref{fig:collapse2} demonstrating the controlling influence of $\phi_m$ in even polydisperse systems.

\subsubsection{\label{sec:level3}Statistically equivalent bidisperse systems}

\begin{figure}[!t] 
%\captionsetup{singlelinecheck=false, justification=centering}
 a) \begin{minipage}[b]{0.33\linewidth}
    \centering
    \includegraphics[width=1.15\linewidth]{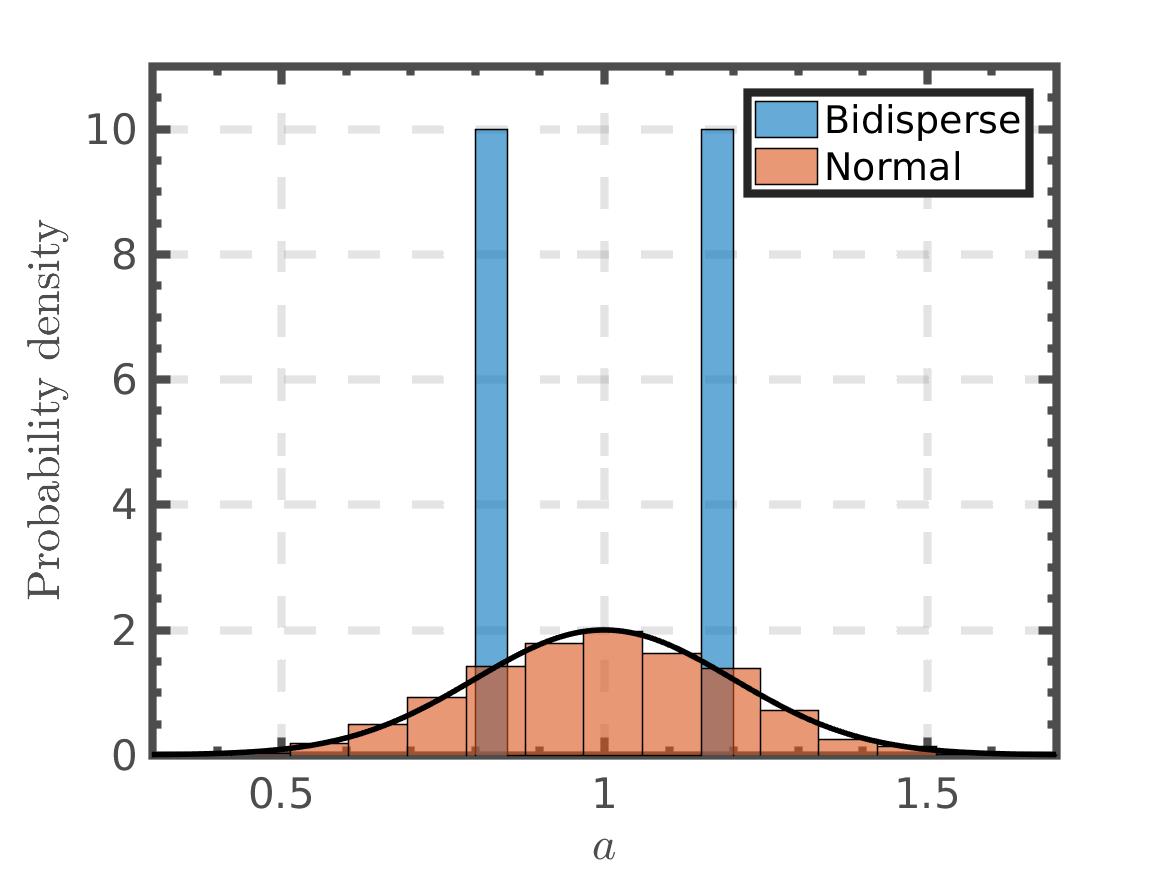} 
    %\caption{DFT, Initial condition} 
   % \vspace{4ex}
  \end{minipage}%% 
  b) \begin{minipage}[b]{0.33\linewidth}
    \centering
    \includegraphics[width=1.15\linewidth]{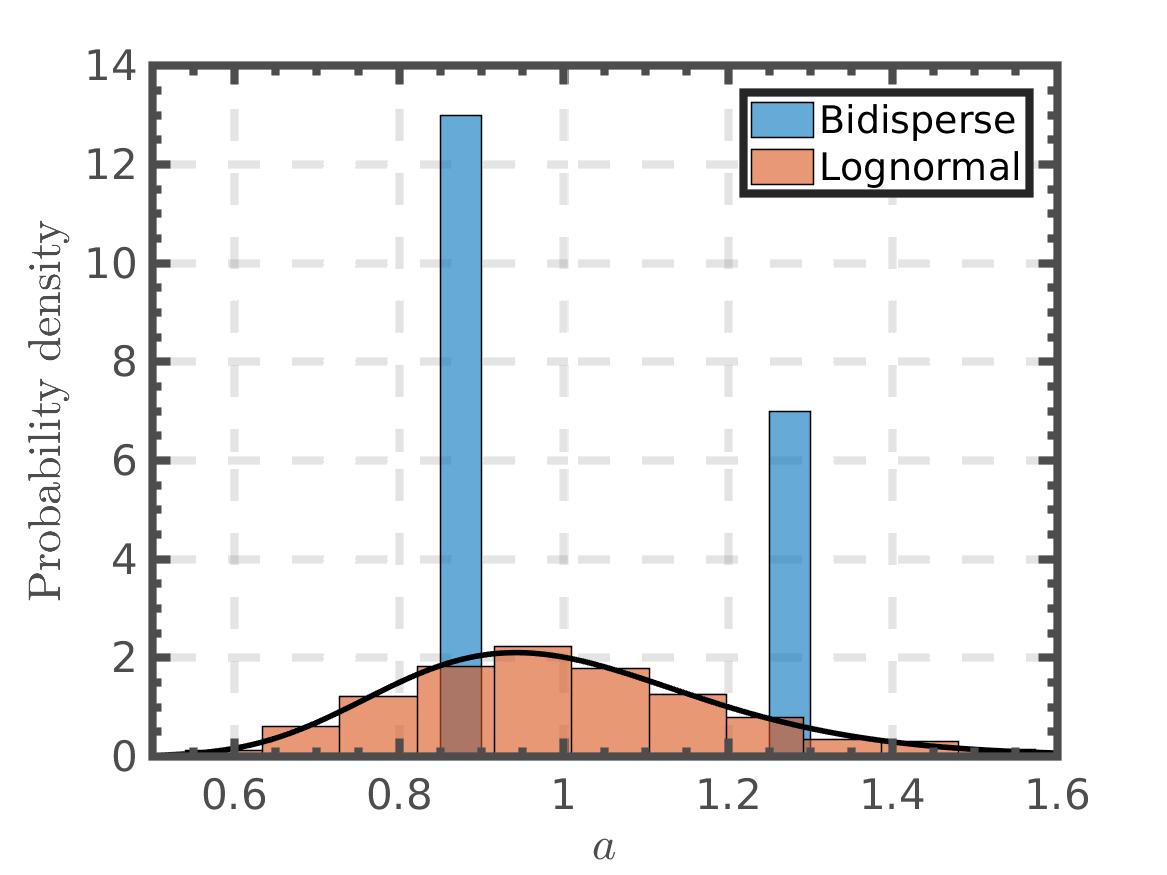} 
\end{minipage} 
    \caption{Histograms of a) normal distribution and b) log-normal distribution with $\alpha$ = 0.2 plotted with those of the bidisperse systems considered `statistically equivalent' (Table \ref{table2}). Black lines represent the analytical form of the probability density for comparison with the generated polydisperse distributions.}% (Fig. \ref from Table {fgr:polypdf})} 
\label{fgr:fgr} 
\end{figure}

In addition to investigating the effects of polydispersity, this work aims to demonstrate rheological equivalence of the relatively complex polydisperse systems with simpler bidisperse systems. The link to this connection as discussed in Sec. \ref{sec:bidi}, lies in having systems of similar maximum packings. Recently, \cite{desmond2014influence} demonstrated that bidisperse and polydisperse systems with equal polydispersity and skewness parameters have similar maximum packings. The expressions for polydispersity, skewness and mean for binary systems are given by 
\begin{equation}
\alpha = [(1-\rho)(a_s-1)^2 + \rho(a_l-1)^2]^{1/2}, \label{1}
\end{equation}

\begin{equation}
S = [(1-\rho)(a_s-1)^3 + \rho(a_l-1)^3]/\alpha^3, \label{2}
\end{equation}

\begin{equation}
\langle a \rangle = \rho a_l + (1-\rho)a_s = 1. \label{3}
\end{equation}
 Here, $\rho$ is the number composition of large particles in the bidisperse mixture, $a_l$ the size of large particles and $a_s$ the size of small particles. As in the case for polydisperse systems, the mean radius of the equivalent bidisperse systems is set to $\langle a \rangle =1$.\\

\begin{table}[!t]
\caption{\label{table2}Polydispersity and skewness of normal and log-normal distributions of radii alongside equivalent bidisperse suspensions parameterized by $\delta$ (Eqn. \ref{delta}) and $\zeta$ (Eqn. \ref{zeta}).}
\begin{ruledtabular}
\begin{tabular}{cccc|cccc}
\multicolumn{2}{c}{Normal}&\multicolumn{2}{c|}{Equiv. Bidisperse}&\multicolumn{2}{c}{Log-normal}&\multicolumn{2}{c}{Equiv. Bidisperse}\\
$\alpha$ & $S$ &$\delta$ & $\zeta$ & $\alpha$ & $S$ &$\delta$ & $\zeta$\\
\hline
0.05& 0 & 1.11 & 0.58 &0.05 & 0.15 & 1.11 & 0.54 \\
0.10& 0 & 1.22 & 0.65 &0.10 & 0.30 & 1.22 & 0.57 \\
0.15& 0 & 1.35 & 0.71 &0.15 & 0.45 & 1.35 & 0.61 \\
0.20& 0 & 1.50 & 0.77 &0.20 & 0.61 & 1.50 & 0.65 \\
0.25\footnotemark[1]& 0 & 1.67 & 0.82 &0.25 & 0.77 & 1.65 & 0.68 \\
0.30\footnotemark[1]& 0 & 1.86 & 0.87 &0.30 & 0.93 & 1.82 & 0.71 \\
\end{tabular}
\end{ruledtabular}
\footnotetext[1]{Simulations not run for this case since $a_{max}/a_{min} > 4$.}
\end{table}

\noindent By matching $\alpha$ and $S$ values of the respective polydisperse systems, we simultaneously solve (\ref{1}-\ref{3}) to determine the three unknowns $\rho$, $a_l$ and $a_s$ of the statistically equivalent bidisperse suspension. We tabulate normal and log-normal polydisperse suspensions with $\alpha$ and their associated skewness $S$ in Table \ref{table2}. Tabulated alongside the polydisperse suspensions are their corresponding equivalent bidisperse suspensions in terms of $\delta$ given by (\ref{delta}) and $\zeta$ given by (\ref{zeta}) converted from $\rho$, $a_l$ and $a_s$. The relationship between large particle volume composition ($\zeta$) and number composition ($\rho$) is given by\\

\begin{equation}
\zeta = \frac{\rho \, \delta^3}{(1-\rho+\rho \, \delta^3)}.
\label{44}
\end{equation}

Fig. \ref{fgr:fgr}a-b illustrate histograms of the generated polydisperse distributions ($\alpha$ = 0.2), plotted along with the corresponding statistically equivalent bidisperse suspensions [Fig. \ref{fgr:fgr}c-d].

\subsubsection{\label{sec:level3}Rheology}

\textit{Effect of polydispersity}:  Fig. \ref{fgr:poly} illustrates the effect of polydispersity at fixed $\phi$ on the rheological response for suspensions of normal (left-side plots) and log normal (right-side plots) distributions of particle size. In Fig. \ref{fgr:poly}a-b we plot the simulated relative viscosity dependence on the polydispersity parameter $\alpha$ for different $\phi$. The first important observation in both normal and log-normal suspensions is that for $\alpha \leq 0.1$, we see little effect of polydispersity on suspension viscosity, with $\eta_r$ values in this range within error bar of each other. This is true also for normal stress responses plotted in Fig. \ref{fgr:poly}c-h. Suspensions having polydispersity $\alpha$ $\leq$ 0.1 can hence seemingly be treated as  monodisperse suspensions. This has been shown true for equilibrium properties in the past [\cite{rastogi1996microstructure}]. For $\alpha > 0.1$, $\eta_r$ decreases with increasing $\alpha$ for both normal and log-normal suspensions. Recall that $\alpha$ is a measure of the spread of the particle size distribution, showing that a greater width in size distribution lowers the viscosity. The effects of the size distribution are, as with bidisperse suspensions [see Sec. \ref{sec:bidi}] more pronounced at higher $\phi$. For the largest $\phi$ = 0.6 (with $>$ 96$\%$ of particles radii falling in the range 1.6 $\geq$ a/$\langle$a$\rangle$ $\geq$ 0.4 for the polydisperse cases), we observe approximately 50\% reduction in viscosity from monodisperse values in normal distributions ($\alpha$ = 0.2) and around 75\% reduction in log-normal systems ($\alpha = 0.3$). 
For similar $a_{max}/a_{min}$, log-normal suspensions have a lower $\eta_r$ than normally distributed suspensions. We can compare these magnitudes of reduction in relative viscosity with reductions due to bidispersity at the same $\phi$ and similar $a_{max}/a_{min}$ (Fig. \ref{fig:eta_delta} and Fig. \ref{fgr:bi}a. We observe that bidisperse suspensions, depending on their volume composition ($\zeta$), can be more effective at reducing relative viscosity. \\
\begin{figure} 
%\captionsetup{singlelinecheck=false, justification=centering}
  a) \begin{minipage}[b]{0.42\linewidth}
    \centering
    \includegraphics[width=1\linewidth]{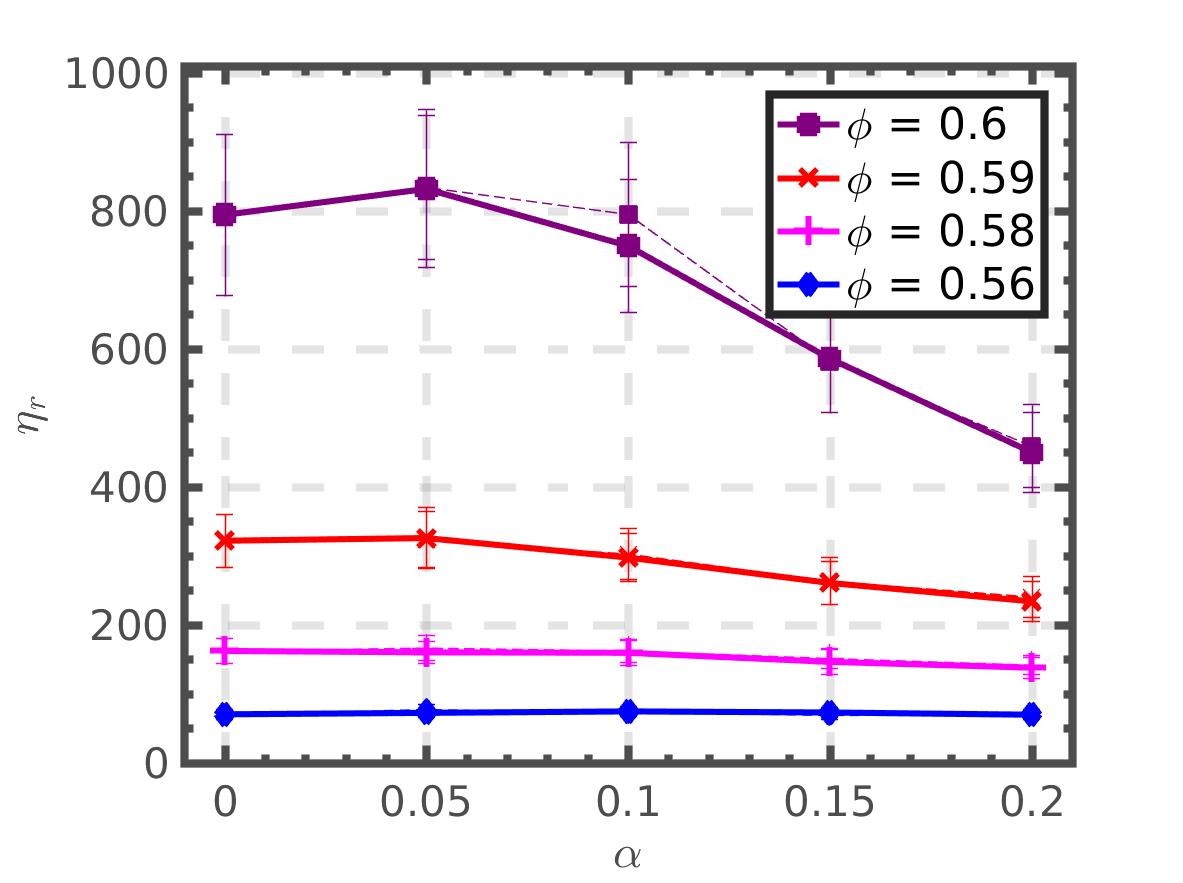} 
    %\caption{Initial condition} 
    %\vspace{4ex}
  \end{minipage}%%
  b) \begin{minipage}[b]{0.42\linewidth}
    \centering
    \includegraphics[width=1\linewidth]{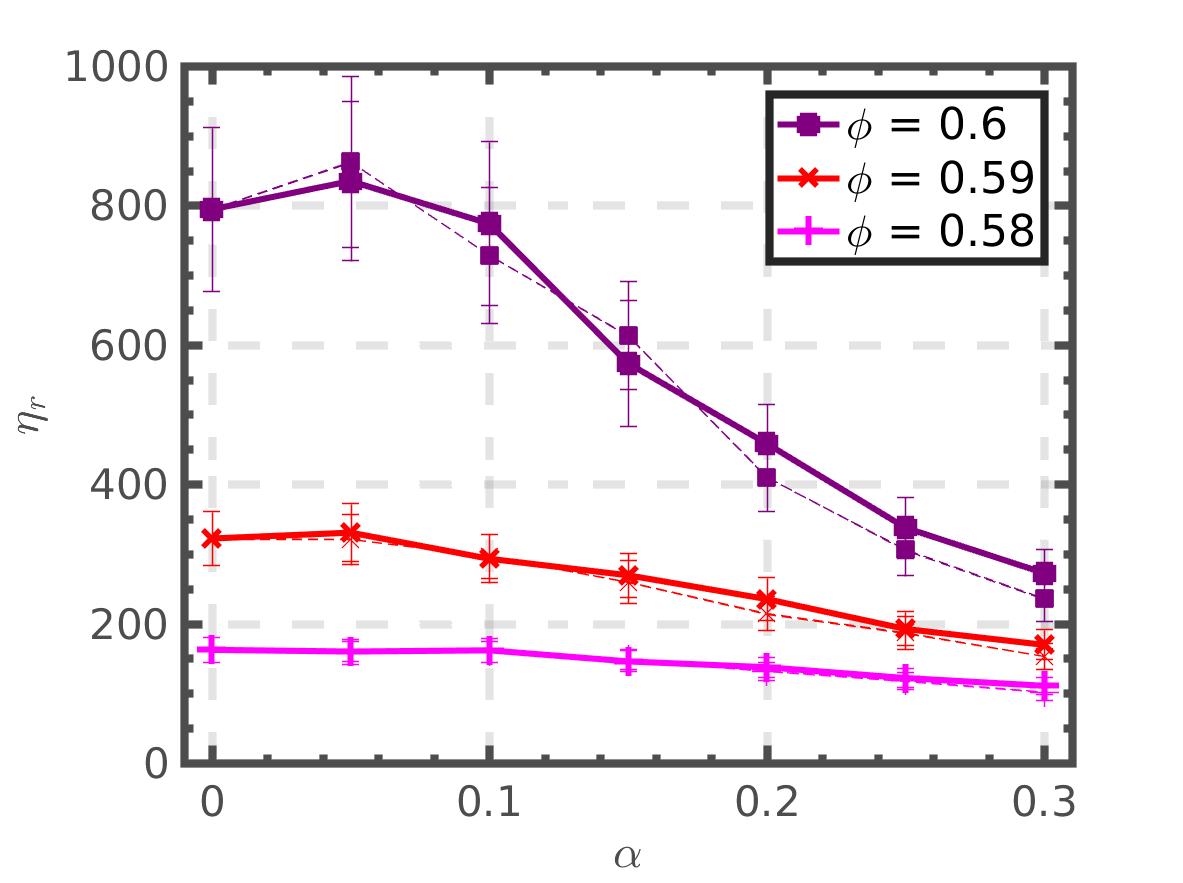} 
    %\caption{Rupture} 
    %\vspace{4ex}
  \end{minipage}\par 
 c) \begin{minipage}[b]{0.42\linewidth}
    \centering
    \includegraphics[width=1\linewidth]{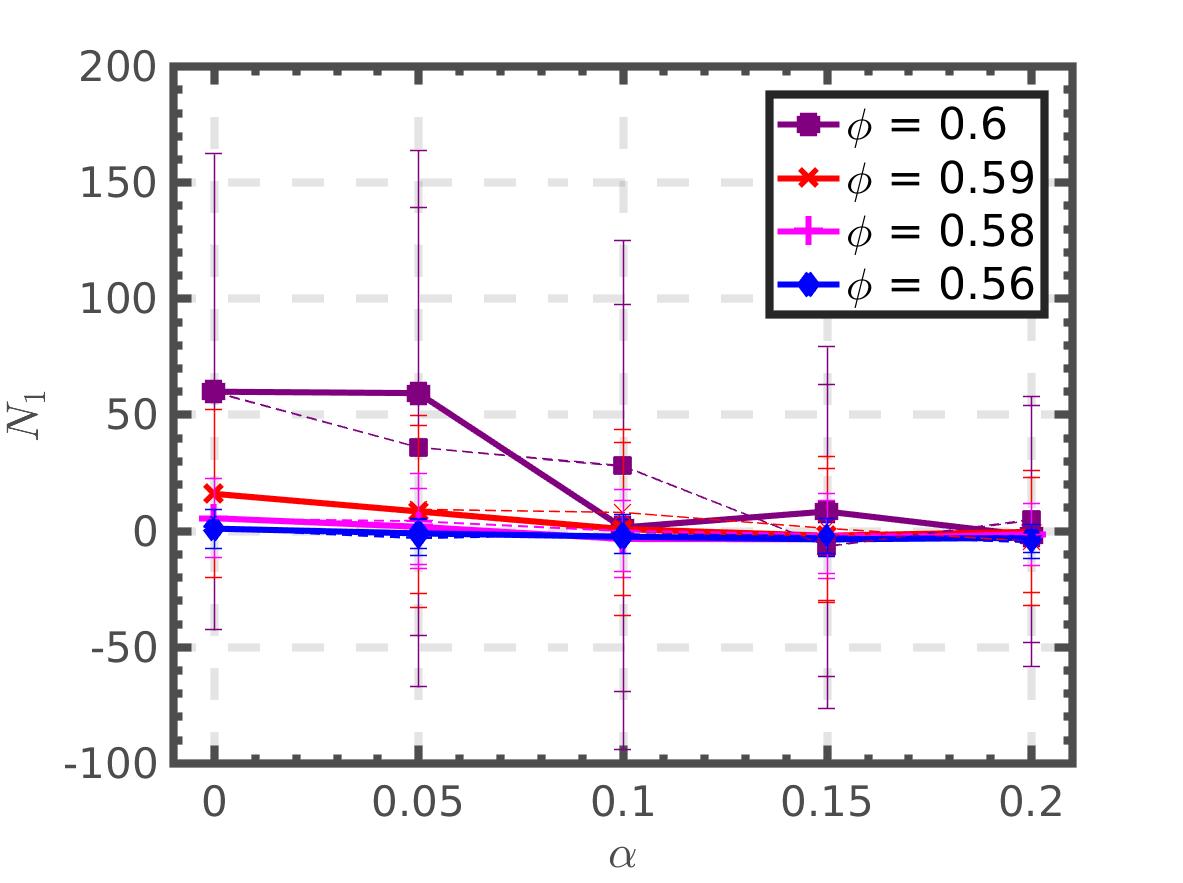}
     %\vspace{4ex}
    %\caption{DFT, Initial condition} 
   % \vspace{4ex}
  \end{minipage}%% 
  d) \begin{minipage}[b]{0.42\linewidth}
    \centering
    \includegraphics[width=1\linewidth]{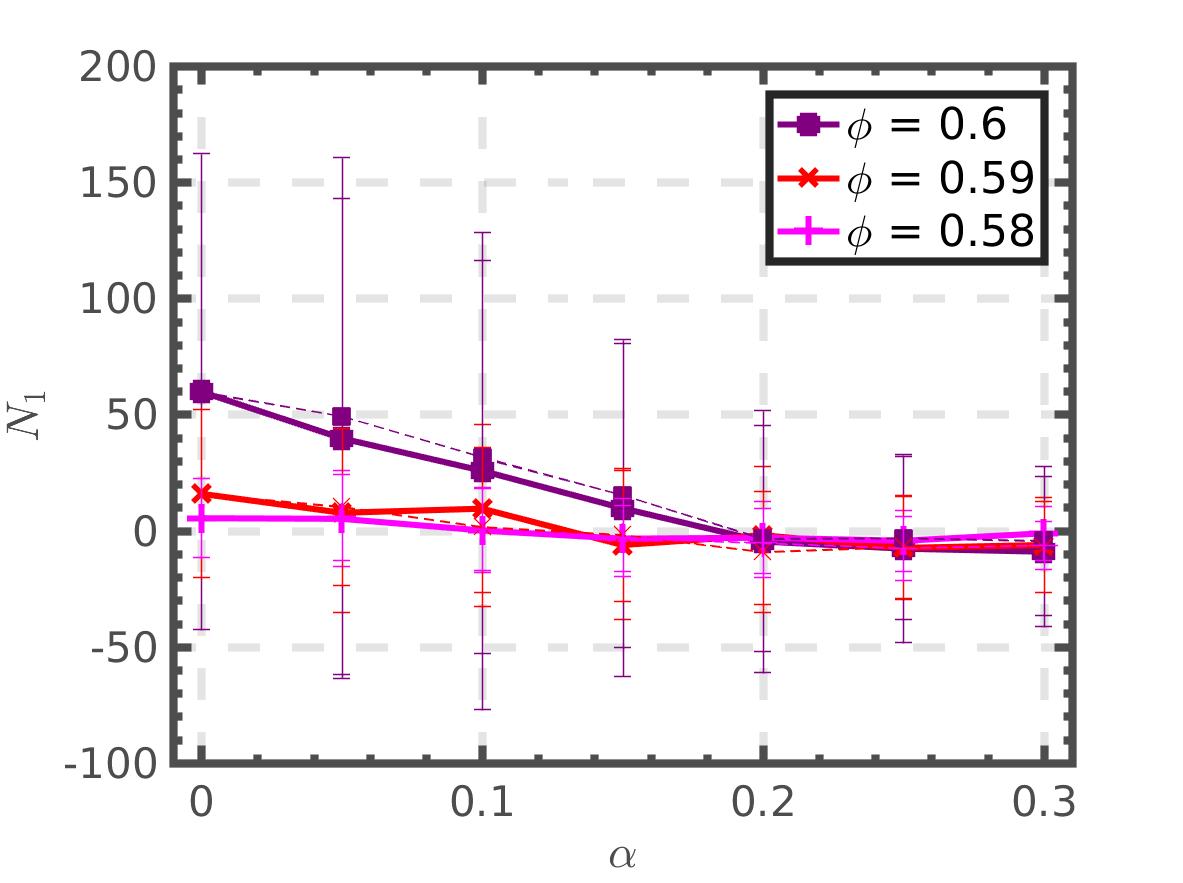} 
     %\vspace{4ex}
    %\caption{DFT, rupture} 
    %\vspace{4ex}
  \end{minipage}

e) \begin{minipage}[b]{0.42\linewidth}
    \centering
    \includegraphics[width=1\linewidth]{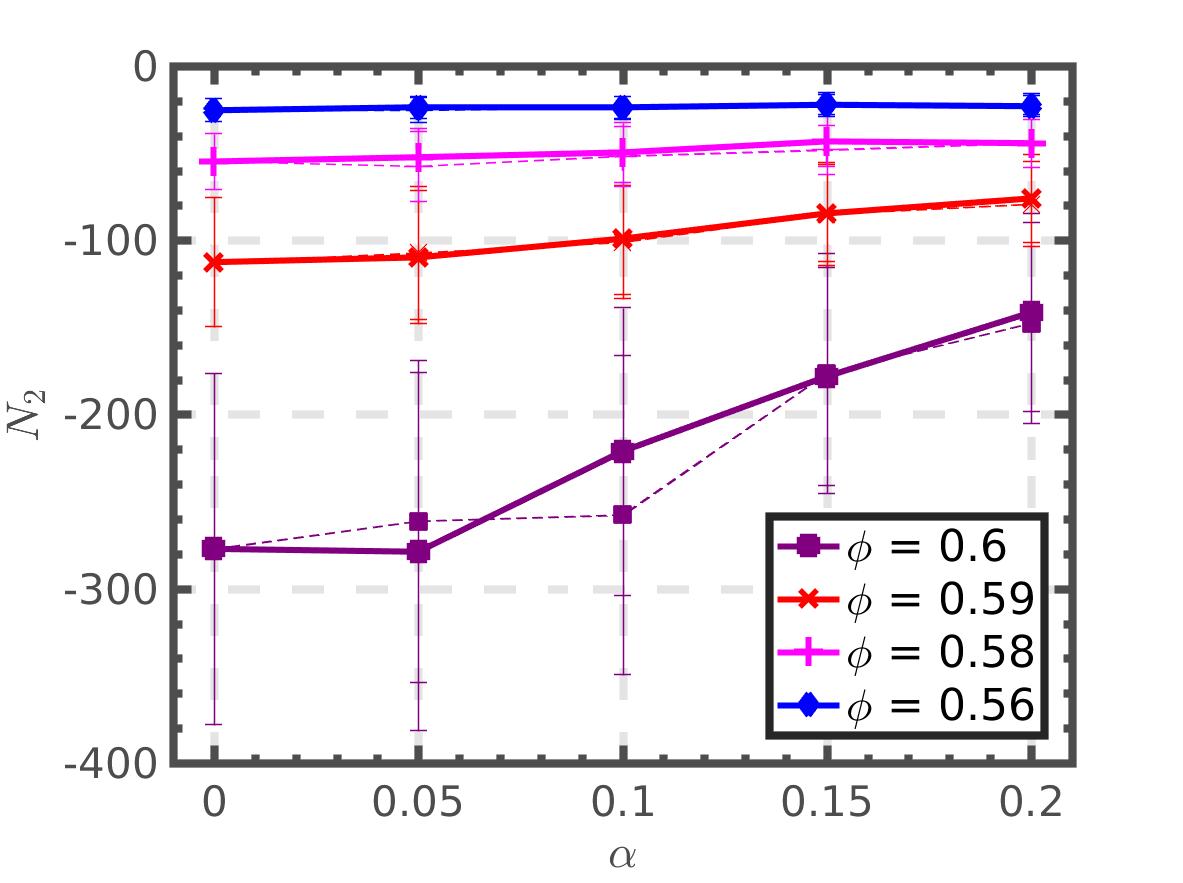} 
    %\caption{Initial condition} 
    %\vspace{4ex}
  \end{minipage}%%
  f) \begin{minipage}[b]{0.42\linewidth}
    \centering
    \includegraphics[width=1\linewidth]{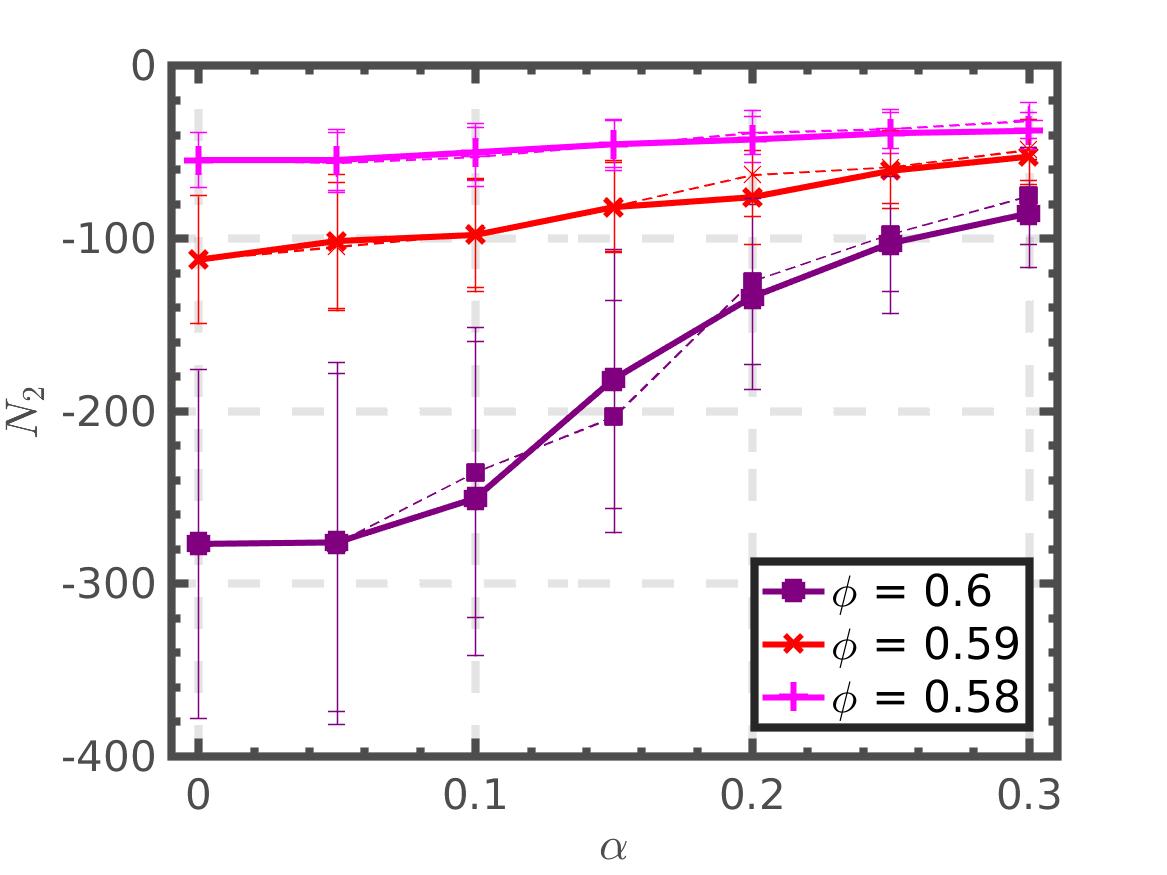} 
    %\caption{Rupture} 
    %\vspace{4ex}
  \end{minipage}\par 
 g) \begin{minipage}[b]{0.42\linewidth}
    \centering
    \includegraphics[width=1\linewidth]{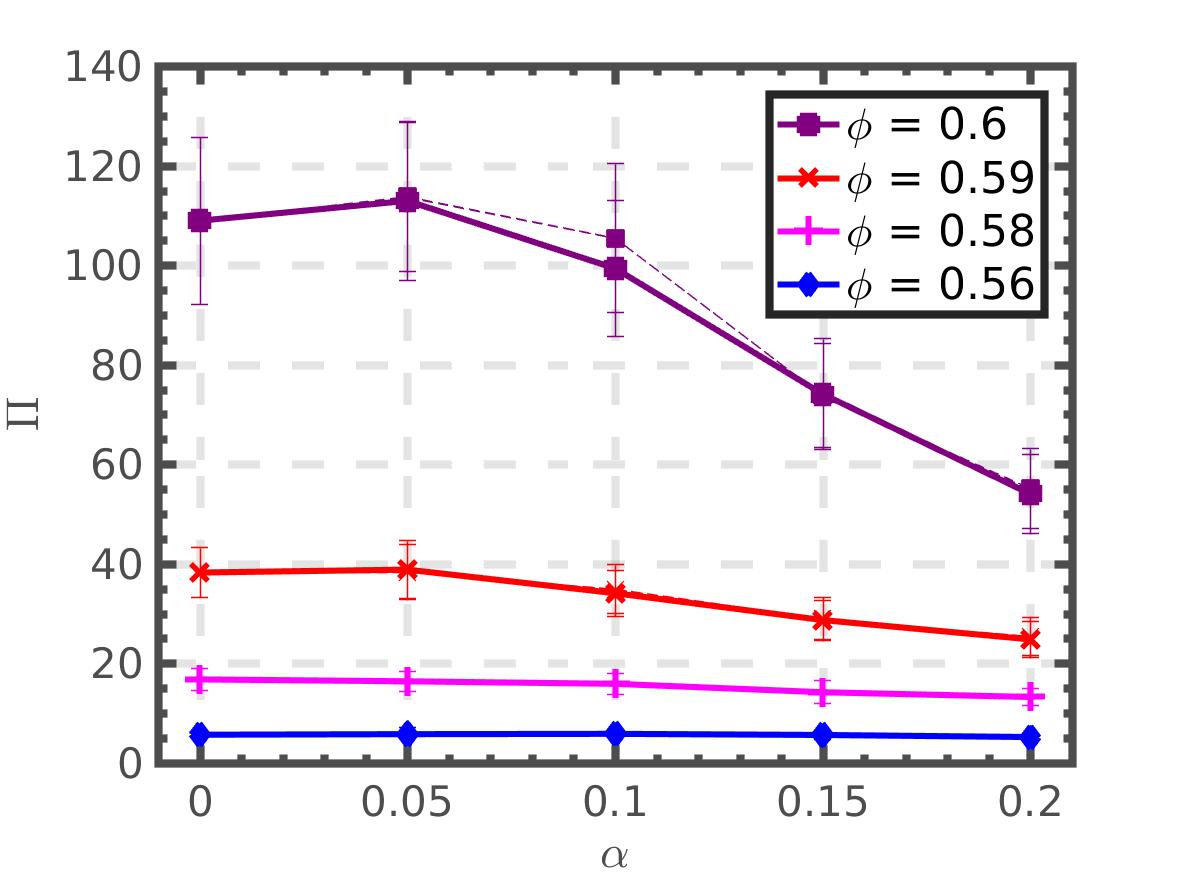}
     %\vspace{4ex}
    %\caption{DFT, Initial condition} 
   % \vspace{4ex}
  \end{minipage}%% 
  h) \begin{minipage}[b]{0.42\linewidth}
    \centering
    \includegraphics[width=1\linewidth]{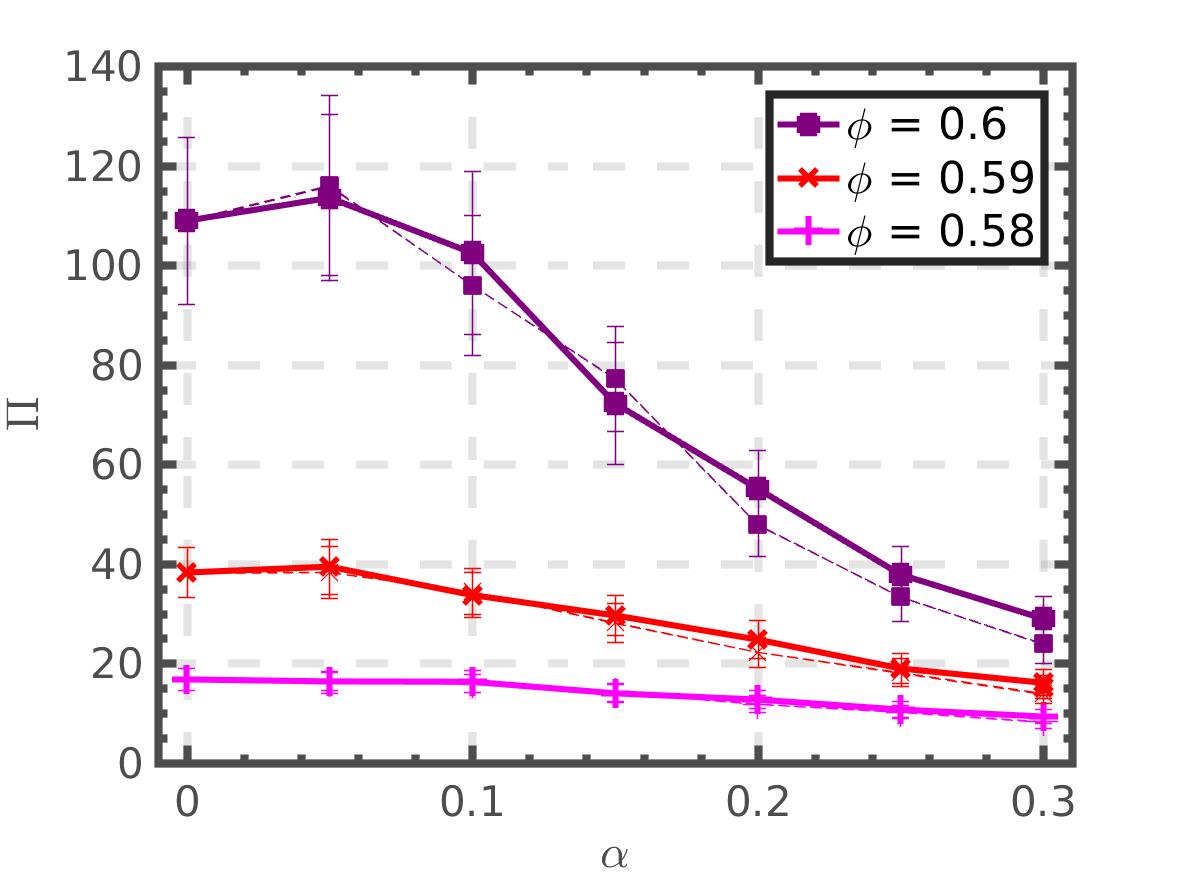} 
  \end{minipage}
  \caption{Rheology of normal (at left) and log-normal suspensions (at right) shown as solid curves, plotted together with `statistically equivalent' bidisperse rheology (dashed lines) a - b) $\eta_r$, c - d) $N_1$, e - f) $N_2$, and g - h) $\Pi$. } 
\label{fgr:poly} 
\end{figure} 

In Fig. \ref{fgr:poly}c-d we observe that the magnitude of the first normal stress difference $N_1$ decreases with $\alpha$. This trend must, however, be interpreted with care given the large fluctuations in $N_1$. Negative $N_2$ and particle pressure $\Pi$ in Fig. \ref{fgr:poly}e-h show similar trends of decrease in magnitude with $\alpha$ as that of $\eta_r$.\\

\textit{Rheology of equivalent bidisperse systems}: Next we compare the rheology of the polydisperse suspensions discussed above with the corresponding `statistically equivalent' bidisperse systems (Table \ref{table2}). As discussed before, the bidisperse suspensions have similar maximum packings as the polydisperse suspensions. In Fig. \ref{fgr:poly}a-f we have plotted the rheology of the equivalent bidisperse suspensions alongside the corresponding polydisperse suspensions. In Fig. \ref{fgr:poly}a-b we observe excellent agreement for relative viscosity between the two  for all volume fractions studied. We emphasize that this equivalence extends to normal stress differences and particle pressure (Fig. \ref{fgr:poly}c-h). \cite{poslinski1988rheological} demonstrated that scaling of first normal stress difference with $\phi$ can be modeled by (\ref{one}) and thus by a primary dependence on $\phi/\phi_m$. Recent work from \cite{singh2017constitutive} has successfully modeled relative
viscosity, $N_2$ and $\Pi$ in shear thickening simulations by considering the jamming point or $\phi_m$ between the low and high viscosity states. \\

\noindent  The equivalent bidisperse suspensions provide a framework for understanding the role of polydispersity on suspension rheology. For instance, as seen in Table \ref{table2}, increasing polydispersity ($\alpha$) is equivalent to increasing bidisperse size ratios ($\delta$). Subtle differences between log-normal and normal distributions can also be gauged from the table using bidisperse rheology as reference (Sec. \ref{sec:bidi}). Although these simulations have focused on high volume fractions, the collapse of relative viscosity curves with $\phi/\phi_m$ [Fig. \ref{fgr:collapse}] suggests that the equivalence may hold true at considerably lower volume fractions; however, testing this would require faithful simulation of long-range hydrodynamics ignored in this work but expected to become relatively important at small volume fraction. It is also likely that small differences in $\phi_m$ from the predicted values will, as $\phi \rightarrow \phi_m$, lead to significant differences between suspensions which are  `equivalent' within the present analysis.\\

\begin{figure}[b]
%\captionsetup{singlelinecheck=false, justification=centering}
  a) \begin{minipage}[b]{0.46\linewidth}
    \centering
    \includegraphics[width=1.15\linewidth]{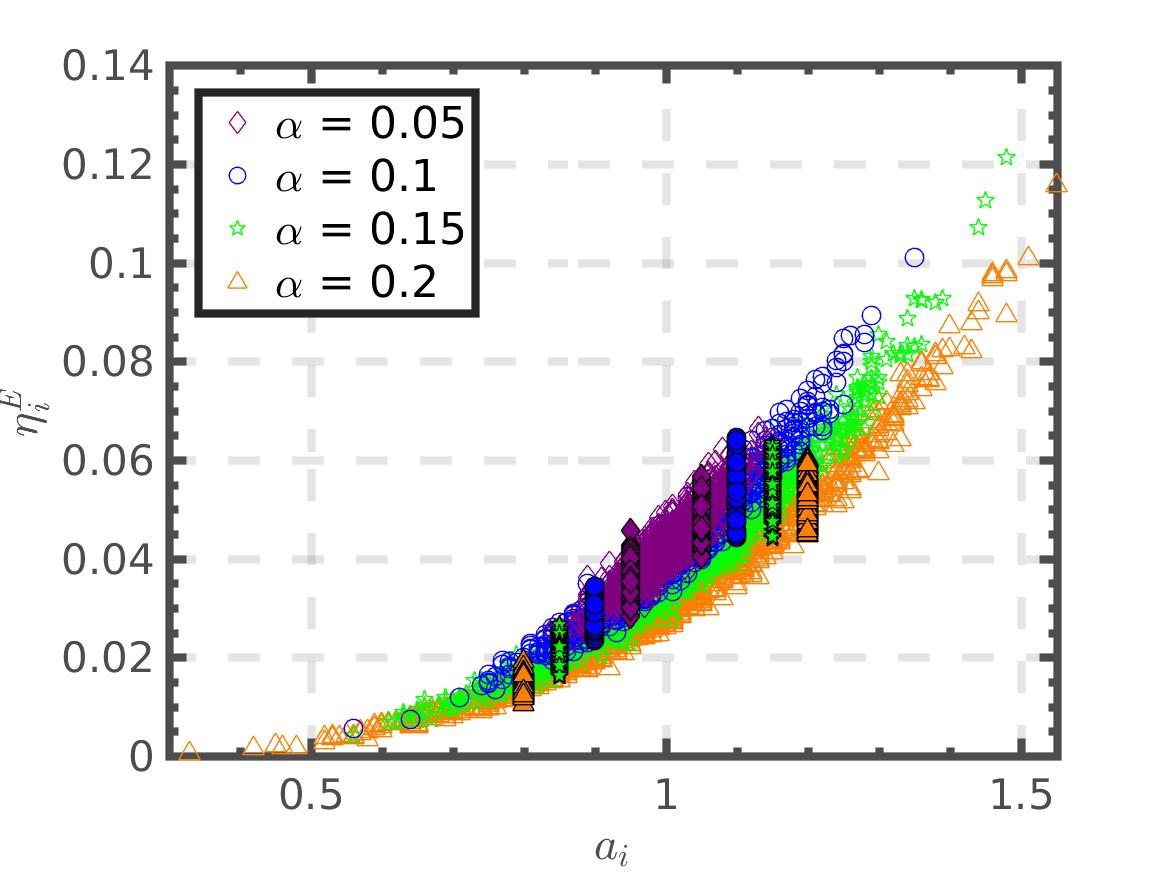} 
    %\caption{Initial condition} 
    %\vspace{4ex}
  \end{minipage}%%
  b) \begin{minipage}[b]{0.46\linewidth}
    \centering
    \includegraphics[width=1.15\linewidth]{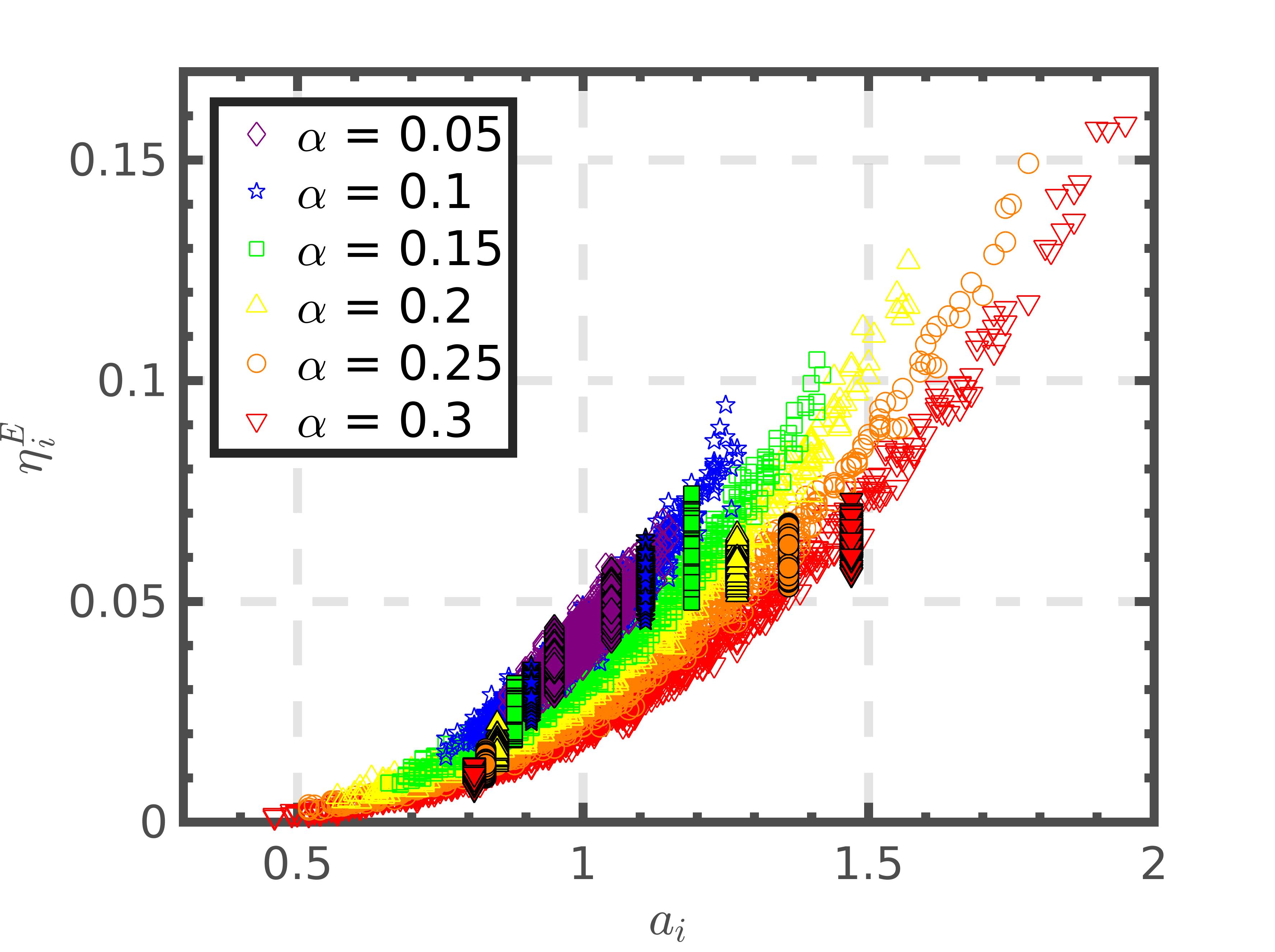} 
    %\caption{Rupture} 
    %\vspace{4ex}
  \end{minipage}\par 
  \caption{Hydrodynamic viscosity contribution as a function of particle radius for each particle for a) normal and b) log-normal suspensions with equivalent bidisperse suspensions at $\phi$ = 0.6.} 
\label{fgr:dist} 
  \end{figure}
  
\textit{Stress as a function of particle size}: Simulation allows us to extract the stress contribution of each particle, an analysis tool that is not readily extended to experiments. In Fig. \ref{fgr:dist}, we plot the hydrodynamic stress exerted by each particle averaged along the length of the simulation and plotted as a function of particle size for both normal (Fig. \ref{fgr:dist}a) and log-normal (Fig. \ref{fgr:dist}b) suspensions.  Each color (or symbol) represents a suspension of a particular polydispersity. The stress distributions with size of these flowing suspensions are seen to be continuous lines. Particles of a given size have higher stresses in suspensions with lower polydispersity. While this is true for  suspensions with $\alpha > 0.1$, polydispersity is seen to have minimal effect for suspensions of lower $\alpha$. This is in agreement with the bulk rheology observations of Fig. \ref{fgr:poly}a-b. The time-averaged hydrodynamic stress as a function of particle size for the equivalent bidisperse suspensions is also plotted  in Fig. \ref{fgr:dist}a-b.  The polydisperse stress curves pass through the corresponding bidisperse lines of the same polydispersity: this says simply that particles of a given size in a polydisperse suspension and its equivalent bidisperse system generate similar levels of stress.

\section{CONCLUSION}
Polydisperse suspensions are more commonly encountered in applications and natural environments than are monodisperse or well-defined bimodal suspensions.  To promote understanding of the role of particle size distribution on rheology, we have systematically investigated by numerical simulation the effect of polydispersity for dense suspensions in the high shear limit. We have shown that it is possible to define a rheologically equivalent bidisperse suspension by matching appropriate moments of the size distribution.   The ability shown in previous works [\cite{chong1971rheology,chang1994effect,stickel2005fluid}] to obtain a collapse of bidisperse suspension viscosity when plotted against reduced volume fraction $\phi/\phi_m$ is shown here to extend to polydisperse suspensions. This approach allows for the modeling of bidisperse and polydisperse suspension viscosity using traditional monodisperse correlations, reducing the problem to determination of the maximum packing fraction.  Recently, \cite{desmond2014influence} showed that the statistical matching of the first three moments (mean, polydispersity and skewness) between bidisperse and different polydisperse packings lead to very similar maximum packings, and thus the information on $\phi_m$ can, in principle, be deduced from the size distribution directly.    In this work, we have tested this by using the approach of \cite{desmond2014influence}  to define rheologically equivalent bidisperse and polydisperse suspensions. Note that an analogous approach has been successfully implemented in studying the rheology of polydisperse granular powders [\cite{gu2016rheology}].  \\

\noindent Examination of the effect of bi- and polydispersity together presents several new insights. The reduction of viscosity in bidisperse suspensions agrees with previous works, and is shown to extend to normal stress differences ($N_1$, $N_2$) and particle pressure.  In polydisperse suspensions, we observe a decrease in suspension viscosity beyond a polydispersity index of $\alpha \approx 0.1$ with more pronounced decrease at higher $\phi$. We propose that polydisperse suspensions with $\alpha<0.1$ can, in rheological terms, be viewed as monodisperse suspensions. A set of criteria for finding rheologically equivalent bidisperse suspensions for polydisperse suspensions is proposed. Simulation results demonstrate that variation of viscosity, $N_1$, $N_2$ and particle pressure of polydisperse suspensions match both qualitatively and quantitatively with the equivalent bidisperse suspension (Table \ref{table2}) of similar $\phi_m$. 
At the microscale, the stress environment has been characterized by determining the particle stress contribution as a function of particle size.   Particles of a given size are found to have a similar stress environment in the polydisperse and equivalent bidisperse suspensions.
 
 While this framework is shown to be successful  for suspensions devoid of inter-particle forces, this may not be true for colloidal particles or particles with anisotropy. 
 Further investigation of polydispersity effects in conjunction with differing shapes and interparticle forces is hence a direction for future work.

\begin{acknowledgments}
The authors gratefully acknowledge discussions with Drs. Hamed Haddadi and Romain Mari. Simulations and manuscript preparation were supported by the Nuclear Process Science Initiative (NPSI), a Laboratory Directed Research and Development (LDRD) Program at Pacific Northwest National Laboratory (PNNL). The computational resources for this research were supported, in part, under National Science Foundation Grants CNS-0958379, CNS-0855217, ACI-1126113 and the City University of New York High Performance Computing Center at the College of Staten Island. Data interpretation and manuscript preparation were supported by the Interfacial Dynamics in Radioactive Environments and Materials (IDREAM), an Energy Frontier Research Center funded by the U.S. Department of Energy (DOE), Office of Science, Basic Energy Sciences. PNNL is a multi-program national laboratory operated for DOE by Battelle under Contract No. DE-AC05-76RL01830.
\end{acknowledgments}

%\nocite{*}
\bibliography{sorsamp}% Produces the bibliography via BibTeX.

\end{document}